\documentclass[prb, 
 reprint,
 amsmath,amssymb,
 aps,
prb,
]{revtex4-2}

\usepackage{graphicx}
\usepackage{dcolumn}
\usepackage{bm}
\usepackage{amsmath}
\usepackage{amssymb}
\usepackage{mathrsfs}
\usepackage[compat=1.1.0]{tikz-feynman}
\usepackage{caption}
\usepackage{feynmp-auto}
\usepackage{tkz-euclide}
\usetikzlibrary{decorations.pathmorphing}
\tikzset{snake it/.style={decorate,decoration=snake}}
\usetikzlibrary{shapes.geometric}

\newcommand{\p}{{\bm \pi}}
\newcommand{\bp}{\bar{{\bm \pi}}}
\def\sumslashD{\mathop{\sum \kern-1.4em -\kern 0.5em}}
\def\sumslash{\mathop{\sum \kern-1.2em -\kern 0.5em}}
\def\intslash{\mathop{\int \kern-0.9em -\kern 0.5em}}
\def\intslashD{\mathop{\int \kern-1.1em -\kern 0.5em}}
\begin{document}

\preprint{APS/123-QED}

\title{Competing charge density-wave and superconducting states in a quasi-one dimensional two-band electron-phonon  model }

\author{C. Bourbonnais}
\email{Claude.Bourbonnais@Usherbrooke.ca}
\author{A. Ghosh}%
 \author{M. Haguier}
\affiliation{Regroupement Québécois sur les Matériaux de Pointe and Institut Quantique,
Département de physique, Université de Sherbrooke, Sherbrooke, Québec, Canada, J1K-2R1
 \\
}%

\date{\today}

\begin{abstract}
We apply the renormalization group method to examine the interplay between the charge density-wave and superconductivity phases in two quasi-one-dimensional tight-binding    electronic bands coupled by the electron-phonon interaction. The electronic and phonon energy scales are embedded in a two-cutoff  scaling scheme at finite temperature for the renormalization group flow equations of  intra- and inter-band phonon-mediated interactions between electrons. At the one-loop level different regimes of renormalization are derived   covering both the adiabatic and  antiadiabatic domains for small and large phonon frequencies. Finite temperature phase diagrams  as a function of nesting frustration of the electron spectrum are thus obtained. The possibilities for superimposed  charge density-wave and superconducting ordered states are given. The emergence of domes of superconductivity resulting from quantum critical reinforcement of Cooper pairing by charge density-wave order fluctuations is found. The impact of the multiband character of the electronic structure on the ordering temperature of superconductivity, quantum criticality, and isotope effect is discussed.  

\end{abstract}

\maketitle


\section{Introduction}
The development of superconductivity in the close vicinity of a charge density-wave (CDW)-ordered state has aroused sustained attention in the last decades as to the possible mechanisms that can condition this proximity. This is well expressed in various materials, such as quasi-one-dimensional organic \cite{Kaddour14,Brossard89,Graf09} and  transition metal trichalcogenides compounds \cite{Monceau12,Briggs80}, layered quasi-two-dimensional transition metal dichalcogenides \cite{Berthier76,Feng12,Tymoshenko25}, Kagome metals \cite{Yu21,Neupert21}, rare earth tellurides \cite{Fang20} and  high-$T_c$ cuprate superconductors \cite{Hayden24}.  Among the characteristics shared by most of these compounds, when, for example, pressure is applied,  the multiplicity of CDW states \cite{Pouget24} is found close to superconductivity (SC). The pattern often develops under quantum critical conditions caused by a puzzling interplay of SC with CDW, in a way apparently distinct from the development of SC close to the  spin density-wave state found, for example, in organics \cite{DoironLeyraud09}. 

Among the common marks underlying the patterns of CDW-SC competing states in these systems is its occurrence in a multi-component electronic structure, which can  come either from distinct but yet coupled pieces of a Fermi surface or else from a multiple Fermi surface resulting from different bands crossing the Fermi level. Such a configuration favors the existence of a CDW-SC manifold and may be an important factor in the formation of quantum criticality and the anomalous increase of the attraction between electrons, and therefore the critical temperature for superconductivity. 
The first theoretical studies to grapple with such intertwined orders  started long ago in the seventies with the mean-field approach to the interplay between the Peierls-like martensitic transition and  phonon-mediated superconductivity in intermetallic compounds \cite{Bilbro76} \cite{Lopes21}, then applied to layered dichalcogenides \cite{Balseiro79,Littlewood81,Grest82} and to systems where  multiple order parameters allowed by symmetry can coexist \cite{Tsonis08}.  Another route has been {\it ab initio} approaches to  the electronic structure and electron-phonon interaction that enter the formation of CDW and conventional  SC  in strong coupling \cite{Nishio94,Heil17}. 

Here, we follow a different path to address  the mechanisms behind  competing CDW-SC orders. This is achieved within a multi-band  electronic structure,  assumed to be dominated by electron-phonon interaction. We shall employ the renormalization group (RG) method that allows to unite on equal footing  the phonon-induced electron-electron  and electron-hole pairings, which are at the core of the microscopic interplay between SC and CDW orders. The RG method is carried out in the Wilsonian scheme at finite temperature \cite{Bourbon91,Bourbon03}, with distinct electronic and phonon energy cutoffs  to incorporate the effects of retardation in the presence of  phonon-mediated interactions \cite{Caron84,Grest76}. 

We organize our paper as follows. In Section~\ref{Model} we proceed to the derivation of  our  tight-binding two-band electron-phonon model and express   the partition function in terms of a path integral formulation for electrons with phonon-mediated interactions. We then introduce the RG transformation for the flow of interactions and pair vertex susceptibilities for different bands at the one-loop level and finite temperature. In Section~\ref{Results} we detail the results of the RG equations for different conditions of adiabaticity and electronic structures of the model and give the respective phase diagrams together with delimiting quantum criticality  and isotope effect for superconductivity. We conclude our work in Section~\ref{conclusion}.  
\section{The electron-phonon model}
\label{Model}
We consider electrons moving on a square lattice, composed of a linear array of $N_b$ chains with $N_a$ sites. Each unit cell basis has two similar orbitals $x=1$, $2$, either atomic or molecular. We follow a tight-binding picture \cite{Barisic70}, in which an electron of spin $\sigma$, represented by the creation and annihilation  operators $a^\dagger_{x,i,\sigma}$  and $ a_{x,i,\sigma}$, can hop between two neighboring lattice sites ${\bm r}_i$ and ${\bm r}_j={\bm r}_i+  {\bm \delta}$ in each direction $\delta_{\mu= a,b}$, and for diagonal ($x=x'$) and off-diagonal ($x\ne x'$) orbitals as a result of overlap integrals $t^{xx'}_{ij}$  ($t^{x\ne x'}_{ij}< t^{xx}_{ij} $).   In the quasi-1D electron picture adopted in this work, $t^{xx'}_{ij}$ is highly anisotropic, with a much stronger amplitude along the chain axis $a$ compared to the perpendicular $b$ direction.   Allowing for an elastic displacement ${\bm R}_{i}\to {\bm R}_{i} + {\bm u}_{{\bm r}_i}$ of atoms or molecules with respect to their equilibrium position ${\bm R}_{i}$,  the overlap integrals are modulated following the first order expansion,  $t^{xx'}_{ij}\simeq t^{xx'}_{\mu} +  \lambda^{xx'}_{\mu}\,(u_j- u_{i})$, where $ \mathbf{e} ={\bm R}_{j}- {\bm R}_{i} $ is a lattice vector and $\lambda^{xx'}_{\mu}= \nabla_{\mathrm{e}_\mu} t^{xx'}_\mu\cdot \hat{\bm \nu}$  is the electron-lattice coupling along the direction $\mu=a,b$ for a lattice displacement  ${\bm u}_{i}= \hat{{\bm \nu}}\, u_i$ when a single polarization of unit vector $\hat{\bm \nu}$ is retained.

The two-orbital quasi-1D  electron-lattice Hamiltonian, $H=H_{\mathrm{p}}^0 + H_{\mathrm{e}} $, consists of $H_p^0$ as the harmonic part of the lattice displacements to be recalled explicitly below, whereas in the above tight-binding picture, the electron part and its coupling to the lattice are given by    
\begin{align}
H_{\mathrm{e}}= & H_{\mathrm{e}}^0 + H_{\mathrm{ep}} \cr
  =&  - \sum_{x,x'}\sum_{i,\sigma,\mu} t^{xx'}_{\mu}(a^\dagger_{x,i,\sigma} a_{x',i+\delta_\mu,\sigma} + a^\dagger_{x,i+\delta_\mu,\sigma} a_{x',i,\sigma})\cr
     & \hskip 0.7 cm -\sum_{x,x'}\sum_{i,\sigma,\mu} \lambda^{xx'}_{\mu} (u_{i+\delta_\mu}-u_{i}) \cr & \hskip 1.15 cm \times (a^\dagger_{x,i,\sigma} a_{x',i+\delta_\mu,\sigma} + a^\dagger_{x,i+\delta_\mu,\sigma} a_{x',i,\sigma}).
\end{align}
We first Fourier transform the bare electronic part $H_{\mathrm{e}}^0$ which leads to 
\begin{equation*}
    H_{\mathrm{e}}^0= -\sum_{xx'}\sum_{{\bm k}} t^{xx'}({\bm k})a^\dagger_{x,{\bm k},\sigma} a_{x',{\bm k},\sigma}
\end{equation*}
with the dispersion relation given by \hbox{$ t^{xx'}({\bm k})=\sum_{\mu} 2t^{xx'}_{\mu}\cos k_\mu d_\mu$} as a function of ${\bm k}=(k_a,k_b)$, where $t^{xx'}_{a}$  and $t^{xx'}_{b}$ are the amplitudes of hopping along and perpendicular to the chains ($ t^{xx'}_{b}\ll t^{xx'}_{a}) $. In what follows, we fix the lattice constants $d_a= d_b=1$. Taking $H_{\mathrm{ep}}$ as a perturbation, one can diagonalize $H_{\mathrm{e}}^0$ in the orbital space by the Bogoliubov transformation $ a^{(\dagger)}_{x,{\bm k},\sigma} =\sum_{\alpha} f_{x,\alpha,{\bm k}}\, c^{(\dagger)}_{\alpha,{\bm k},\sigma} $ for the new orbitals $\alpha= \p,\bp$, with the real coefficients $ f_{x,\alpha,{\bm k}}$ and for which $f_{x,\p,{\bm k}}^2+ f_{x,\bp,{\bm k}}^2=1$. 

Expressing the Fourier components $u_{\bm q}$ of the lattice displacement in terms of the phonon operators $b^\dagger_{\bm q}  $ and $b_{\bm q}$ of wave vector ${\bm q}=(q_a,q_b)$ and energy $\omega_{\bm q}$ ($\hbar =1$), the bare phonon Hamiltonian takes its harmonic form $H_{\mathrm{p}}^0=\sum_{\bm q}\omega_{\bm q} (b^\dagger_{\bm q}b_{\bm q}+1/2)$, where the phonon frequency $\omega_{q_\mu} =\omega_D\sin q_\mu/2$ is related to the Debye frequency $\omega_D= \sqrt{2\kappa/M}$ and where $\kappa$ is the string constant  and $M$ the molecular mass. The electron-phonon part, $H_{\mathrm{e}}$, becomes 
\begin{align}
  &H_{\mathrm{e}} =   \sum_{\alpha, \sigma,{\bm k}} \epsilon^\alpha_{\bm k} c^\dagger_{\alpha,{\bm k},\sigma}c_{\alpha,{\bm k},\sigma}\cr
  &\! \! +\! \frac{1}{\sqrt{N_aN_b}}\sum_{\alpha,\delta} \sum_{{\bm k},{\bm q},\sigma}\! g^{\alpha\delta}({\bm k},{\bm q})  c^\dagger_{\alpha,{\bm k}+{\bm q},\sigma}c_{\delta,{\bm k},\sigma}(b^\dagger_{\bm q} + b_{-{\bm q}}),\,
\end{align}
where $\epsilon^{\alpha=\p,\bp}_{\bm k} = \bar{t}({\bm k})\pm \frac{1}{2}\sqrt{(\Delta t({\bm k}))^2+ (2t^{12}({\bm k}))^2}$ , with $ \bar{t}({\bm k}) = \frac{1}{2}(t^{11}({\bm k}) + t^{22}({\bm k})) $  and $\Delta t({\bm k})=  t^{22}({\bm k})- t^{11}({\bm k})$;   and
\begin{align}
 g^{\alpha\delta}({\bm k},{\bm q}) =  -\frac{4 i}{\sqrt{2M\omega_{\bm q}}}& \sum_\mu\sum_{x,x'} f_{x,\alpha,{k}_\mu+{q}_\mu}\lambda^{xx'}_\mu  f_{x,\delta,{k}_\mu} \cr & \times \cos({k}_\mu + {q}_\mu/2)\sin (q_\mu/2),  
 \label{ephv}
\end{align}
as the amplitude of electron-phonon scattering between the bands $\alpha$ and $\delta$.
\subsection{The partition function }
\begin{figure}
    \centering
    \includegraphics[width=1.0\linewidth]{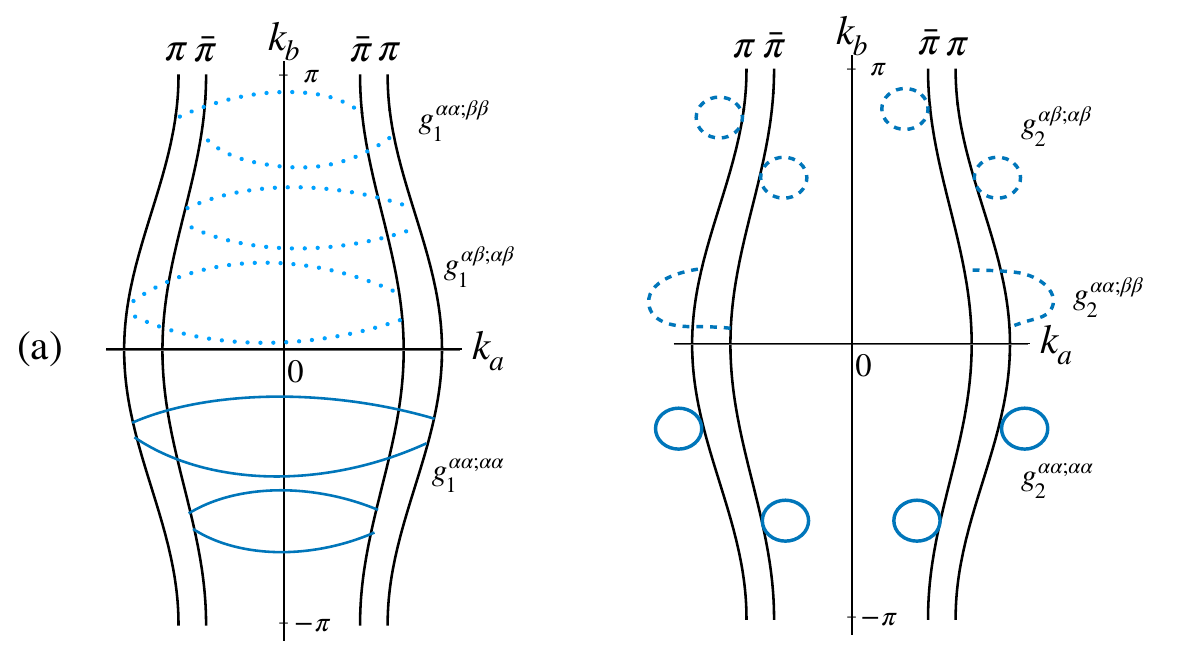}
    \begin{tikzpicture}[scale=0.8]
\node at (-4,0) {(b)};
        \draw[-, thick](-3,1) -- (-1.5,1) node[midway,above=2pt]{$\delta$} ; 
        \node[isosceles triangle, draw, fill = black, minimum size= 0.15cm, inner sep = 0pt, rotate = 0] at (-2.3,1){} ; 
        \draw[dashed, thick](-1.5,1) -- (0,1) node[midway,above=2pt]{$\alpha$} ;
        \node[isosceles triangle, draw, fill = black, minimum size= 0.15cm, inner sep = 0pt, rotate = 0] at (-0.75,1){};
        \filldraw[black](-1.5,1) circle (1.5pt);
        \draw[dashed, thick](-3,-1.5) -- (-1.5,-1.5) node[midway,above=2pt]{$\gamma$}; 
        \node[isosceles triangle, draw, fill = black, minimum size= 0.15cm, inner sep = 0pt, rotate = 0] at (-2.3,-1.5){} ;
        \draw[snake it, thick](-1.5,1) -- (-1.5,-1.5) node[midway,right=3pt]{$g_1^{\alpha\beta;\gamma\delta}$};
        \draw[-, thick](-1.5,-1.5) -- (0,-1.5)  node[midway,above=2pt]{$\beta$} ;
        \node[isosceles triangle, draw, fill = black, minimum size= 0.15cm, inner sep = 0pt, rotate = 0] at (-0.75,-1.5){} ;
        \filldraw[black](-1.5,-1.5) circle (1.5pt);
        \draw[-, thick](1,1) -- (2.5,1) node[midway,above=2pt]{$\delta$} ; 
        \node[isosceles triangle, draw, fill = black, minimum size= 0.15cm, inner sep = 0pt, rotate = 0] at (1.75,1){} ; 
        \filldraw[black](2.5,1) circle (1.5pt);
        \draw[-, thick](2.5,1) -- (4.0,1) node[midway,above=2pt]{$\alpha$} ;
        \node[isosceles triangle, draw, fill = black, minimum size= 0.15cm, inner sep = 0pt, rotate = 0] at (3.25,1){};
        \draw[snake it, thick](2.5,1) -- (2.5,-1.5) node[midway,right=3pt]{$g_2^{\alpha\beta;\gamma\delta}$};
        \filldraw[black](2.5,-1.5) circle (1.5pt);
        \draw[dashed, thick](1,-1.5) -- (2.5,-1.5) node[midway,above=2pt]{$\gamma$}; 
        \node[isosceles triangle, draw, fill = black, minimum size= 0.15cm, inner sep = 0pt, rotate = 0] at (1.75,-1.5){} ;
         \draw[dashed, thick](2.5,-1.5) -- (4.0,-1.5)  node[midway,above=2pt]{$\beta$} ;
        \node[isosceles triangle, draw, fill = black, minimum size= 0.15cm, inner sep = 0pt, rotate = 0] at (3.25,-1.5){} ;
           \end{tikzpicture}
           \caption{(a) Two-band quasi-1D open Fermi surfaces and examples of intra-band (continuous blue) and inter-band (dashed blue)  backscattering (left) and forward (right) scattering; (b) backward ($g_1$) and forward ($g_2$) scattering amplitudes with the band labels. Full (dashed) lines refer to  right (left) going electron along the chain $a$-direction.\label{g1g2}  }
    \end{figure}
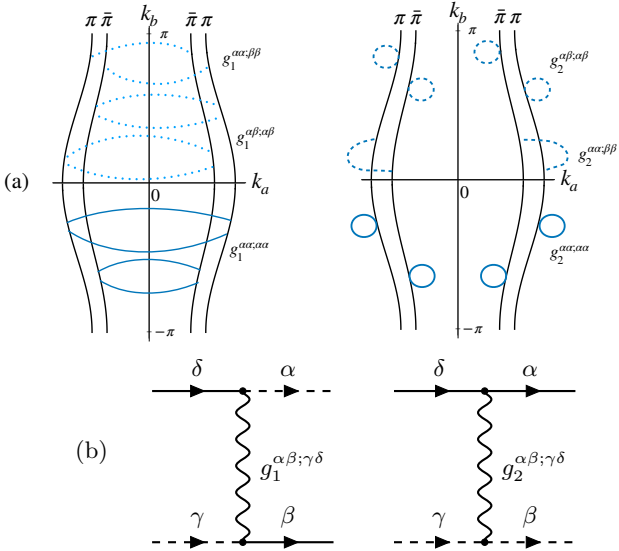

In the interaction picture, the partition function is basically written as
\begin{equation}
    Z=\mathrm{Tr}_{\mathrm{e}}\mathrm{Tr}_{\mathrm{p}} e^{-\beta(H_p^0 + H_e^0- \sum_\alpha \mu_\alpha N_\alpha)}  {\cal T}_\tau\Big\{ e^{-\int_0^{\beta} H_{ep}(\tau)d\tau}\Big\}.
\end{equation}
where $\mu_\alpha$ and $N_\alpha$ are the chemical potential and the number operator for band $\alpha$, and ${\cal T}_\tau$ is the ordering operator along the imaginary time axis $\tau$ ($\beta =1/T$, $k_B=1$). In the standard way, the trace Tr$_{\mathrm{p}}$ over harmonic phonon degrees of freedom can be carried out exactly to generate a $\tau$-retarded phonon-induced interaction between electrons of the bands involved. By expressing the resulting partition function in the functional integral representation, we can write 
\begin{equation}
    Z[h]= \iint \mathcal{D}\psi^*\mathcal{D}\psi\  e^{S}
    \label{Zh}
\end{equation}
where $S= S_0 + S_I + S_h $ is the total action function of the anticommuting fermion fields $\psi(\psi^*)$. In  the Fourier-Matsubara space, it first comprises 
\begin{equation}
    S_0[\psi^*,\psi] = \sum_{\alpha}\sum_{p,k,\sigma}[G_{p}^\alpha(k)]^{-1}\psi^*_{\alpha,p,k}\psi_{\alpha,p,k},
\end{equation}
as the free-electron part. It is defined by the free propagator of the band $\alpha$,
\begin{equation}
    G_p^\alpha(k) = [i\omega_n - E_{p,{\bm k}}^\alpha]^{-1},
\end{equation}
where $k\equiv({\bm k},\omega_n)$ and $\omega_n=\pm\pi T, \pm 3\pi T, \ldots$. The spectrum $E_{p,{\bm k}}^\alpha=\epsilon^\alpha_{\bm k}-\mu_\alpha$ of the band $\alpha$  with respect to  its Fermi level will be approximated by 
\begin{align}
    \ E_{p,{\bm k}}^\alpha \simeq &  \ -2t^\alpha_a\cos k_a -2t_b^\alpha\cos k_b -\mu_\alpha  \cr
   \simeq  &\  v_F^\alpha (pk_a-k_F^\alpha) -2t^\alpha_b\cos k_b -2t'^\alpha_b\cos 2k_b,\cr 
   \label{Epk}
\end{align}
  for relatively small difference   $|\Delta t({\bm k})|/|t^{12}({\bm k})| < 1$ in orbital overlaps, where $t^{\alpha=\p,\bp}_\mu = \frac{1}{2}(t_\mu^{11}+t_\mu^{22} \pm 2 t_\mu^{12})$. In  the second line of (\ref{Epk}), the  part parallel to the chains has been linearized around the open Fermi surface of the band $\alpha$ shown in Figure~\ref{g1g2}-(a), $p k_F^\alpha $, for a moving electron on the right ($p=+$) or left ($p=-$) along the $a$ axis, and where  $k_F^\alpha$ and $v_F^\alpha = 2t_a^\alpha \sin k_F^\alpha$ are the longitudinal Fermi wave vector and velocity, respectively;   $t_b^\alpha$ and  $t'^\alpha_b=-(t_b^\alpha)^2\cos k_F^\alpha/(4t_a^\alpha \sin^2k_F^\alpha)$ are the perpendicular hopping terms between the first and next to the nearest neighbor chains. The latter is a small correction generated by the longitudinal linearization of the spectrum away from half-filling for the bands. The correction plays a significant role in altering the electron-hole symmetry (nesting) \cite{Yamaji82}, by simulating the effect of applied pressure on the Fermi surfaces in concrete situations; it will be taken as a varying parameter in our multiband model. We next introduce 
 \begin{equation}
     \bar{v} \equiv \frac{v_F^{\bp}}{v_F^\p}= \frac{t_b^{\bp}}{t_b^{\p}}= \frac{t_b'^{\bp}}{t_b'^{\p}}
 \end{equation}
 that parameterizes the parallelism between the bands. 

 The part of the action for the phonon-induced interaction between electrons, when decomposed into a g-ology picture, is found to be 
 \begin{align}
     S_I[\psi^*,\psi]= & - \frac{T}{N_aN_b}\sum_{\alpha,\beta,\gamma,\delta} \sum_{k,k',q}\sum_{\sigma\sigma'}\Big\{g_1^{\alpha\beta;\gamma\delta}\mathscr{D}(\omega_m)\cr
     &\times \psi^*_{-,\alpha,k-q,\sigma}\psi^*_{+,\beta,k'+q,\sigma'}\psi_{-,\gamma,k',\sigma'}\psi_{+,\delta,k,\sigma}\cr 
     &  \hskip 3.6 cm + g_2^{\alpha\beta;\gamma\delta}\mathscr{D}(\omega_m)\cr
     &\times \psi^*_{+,\alpha,k+q,\sigma}\psi^*_{-,\beta,k'-q,\sigma'}\psi_{-,\gamma,k',\sigma'}\psi_{+,\delta,k,\sigma}\Big\},
     \cr
     \label{SI}
      \end{align}
      where $g_1^{\alpha\beta;\gamma\delta}$ and $g_2^{\alpha\beta;\gamma\delta}$  are the amplitudes of backscattering  and forward scattering of electrons on opposite longitudinal branches $p$ and for ingoing $(\gamma,\delta)$  and outgoing $(\alpha,\beta) $ bands (see Fig.~\ref{g1g2}). In the g-ology framework, these are taken as momentum-independent coupling constants defined from an average over the quasi-1D Fermi surfaces:
      \begin{align}
          g_{\iota=1,2}^{\alpha\beta;\gamma\delta}= -\frac{2}{\omega_D}\big\langle\, g^{\alpha\delta}&({\bm k}^\delta_{F,+}(k_b),-{\bm q}_{\iota,\alpha\delta})\cr 
          \times& g^{\beta\gamma}({\bm k}^\gamma_{F,-}(k'_b),{\bm q}_{\iota,\alpha\delta})\,\big\rangle_{k_b,k_b'}.
          \label{g12}
      \end{align}
Here ${\bm k}^\alpha_{p,F}(k_b)=(k^\alpha_{p,a}(k_b),k_b)$ is the Fermi wave vector of the branch $p$ and band $\alpha$; the longitudinal component $k^\alpha_{p,a}(k_b)$ is determined from (\ref{Epk}) by the condition $E^\alpha_{p,{\bm k}_F}=0$ on the Fermi surface. The momentum transfer $q_{\iota,\alpha\delta}$ is associated with each coupling, and its possible characteristic intraband or interband values will be listed below. The expression in brackets $\langle \ldots \rangle_{k_b,k_b'} = (2\pi)^{-2} \iint dk_bdk_b' \ldots $ is an average over the Fermi surfaces. The phonon propagator in $S_I$ is written in the dimensionless form
\begin{equation}
   \mathscr{D}(\omega_m) = [1+ (\omega_m/\omega_D)^2]^{-1},
\end{equation}
which will be taken the same for both couplings and where $\omega_D$ is the characteristic phonon frequency.

The last term of the action in (\ref{Zh}) is 
\begin{align}
    S_h[\psi^*,\psi] & =\   h_{\mathrm{CDW}}^{\alpha\gamma}(q_{\alpha\gamma}) z_{\mathrm{CDW}}^{\alpha\gamma}  \cr & \ \times \frac{T}{\sqrt{N_aN_b}}\sum_{p,k,\sigma }\psi^*_{-p,\alpha,{k-pq_{\alpha\gamma}},\sigma}\psi_{p,\gamma,{ k},\sigma}\\
     & \ \  + h_{\mathrm{SC}}^{\alpha\alpha}z_{\mathrm{SC}}^{\alpha\alpha}  \cr & \ \times \frac{T}{\sqrt{N_aN_b}}\sum_{k,\sigma}\big[\sigma \psi_{-,\alpha,-k,\sigma}\psi_{+,\alpha,{ k},\sigma}\cr
     & \hskip 2.5 cm + \sigma\psi^*_{+,\alpha,k,\sigma}\psi^*_{-,\alpha,{ -k},\sigma}\big]
\end{align}
and consists of the addition of two sets of source fields $h$ linearly coupled to the static electronic CDW at the external vector $q_{\alpha\gamma}=({\bm q}_{\alpha\gamma},0)$ and to the static singlet intraband SC superconductivity at zero pair momentum. Here, the modulation wave vector $ {\bm q}_{\alpha\gamma}$ can be for intraband (${\bm q}_{\alpha\alpha}$) or interband (${\bm q}_{\alpha\ne \gamma}$) CDW; they actually coincide with the possible momentum transfer ${\bm q}_{1,\alpha\gamma}$ for backscattering in (\ref{g12}).    The related coefficients $z_{\mathrm{CDW}}^{\alpha\gamma}$ and $z_{\mathrm{SC}}^{\alpha\alpha} $ are in the definition of susceptibilities for CDW and SC given below [See Eqs.~(\ref{KiCDW}) and (\ref{KiSC})].  At the bare level, $z_{\mathrm{CDW}}^{\alpha\gamma}=z_{\mathrm{SC}}^{\alpha\gamma}=1 $. 

\subsection{The renormalization group equations}
\label{RGEqs}
The application of the perturbative renormalization group method for retarded phonon-induced interactions like those in (\ref{SI})  requires at the outset some attention. Our model is characterized by two distinct energy scales, the Fermi energy $E_F^\alpha$ for electrons and the phonon energy $\omega_D$, which in practice are expected to be quite different, namely, $\omega_D\ll E_F^\alpha$. This is signaled in the perturbation theory of coupling constants \cite{Grest76}.

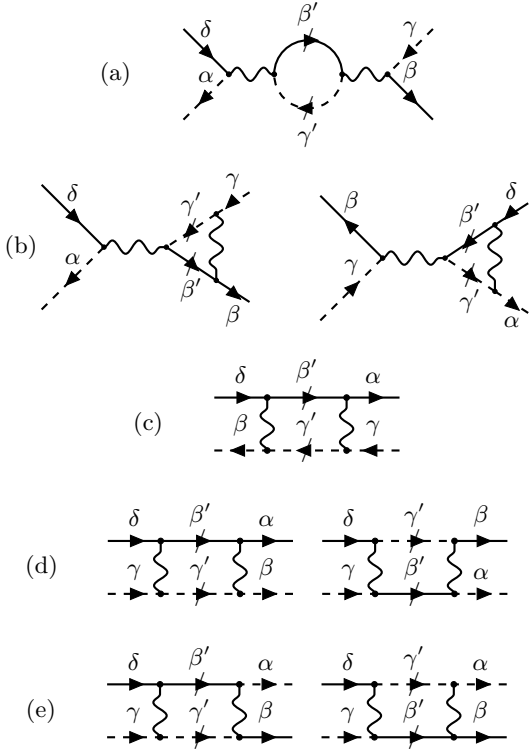
\begin{figure}
    \centering
    \begin{tikzpicture}[scale=0.6]
\node at (-4,0) {(a)};
        \draw[-, thick](-2.5,1) -- (-1.5,0) node[midway,above=2pt]{$\delta$} ; 
        \node[isosceles triangle, draw, fill = black, minimum size= 0.15cm, inner sep = 0pt, rotate = -45] at (-2,0.5){} ;
        \draw[snake it, thick](-1.5,0) -- (-0.5,0); 
        \draw[dashed, thick](-2.5,-1) -- (-1.5,0) node[midway,above=2pt]{$\alpha$} ;
        \node[isosceles triangle, draw, fill = black, minimum size= 0.15cm, inner sep = 0pt, rotate = 225] at (-2,-0.5){} ;
        \filldraw[black](-1.5,0) circle (1.5pt);
        \filldraw[black](-0.5,0) circle (1.5pt); 
        \draw[dashed, thick](3,1) -- (2,0) node[midway,above=2pt]{$\gamma$} ; 
        \node[isosceles triangle, draw, fill = black, minimum size= 0.15cm, inner sep = 0pt, rotate = -45] at (2.5,-0.5){} ;
        \draw[snake it, thick](2,0) -- (1,0);
        \draw[-, thick](3,-1) -- (2,0)  node[midway,above=2pt]{$\beta$} ;
        \node[isosceles triangle, draw, fill = black, minimum size= 0.15cm, inner sep = 0pt, rotate = 225] at (2.5,0.5){} ;
        \filldraw[black](1,0) circle (1.5pt);
        \filldraw[black](2,0) circle (1.5pt);
        \draw[-, thick](1,0) arc (0:180:0.75) node[midway,above=2pt]{$\beta'$} node[midway]{\slash} ;
         \node[isosceles triangle, draw, fill = black, minimum size= 0.15cm, inner sep = 0pt, rotate = 0] at (0.25,0.75){} ;
        \node[isosceles triangle, draw, fill = black, minimum size= 0.15cm, inner sep = 0pt, rotate = 180] at (0.25,-0.75){} ;
        \draw[dashed, thick](1,0) arc (0:-180:0.75) 
        node[midway,below=2pt]{$\gamma'$}node[midway]{\slash};
    \end{tikzpicture}
    \\[3pt]
    \begin{tikzpicture}[scale=0.55]
    \node at (-2,0) {(b)};
        \draw[-, thick](-1.5,1.5) -- (0,0) node[midway,above=2pt] {$\delta$} ; 
        \node[isosceles triangle, draw, fill = black, minimum size= 0.15cm, inner sep = 0pt, rotate = -45] at (-0.75,0.75){}  ;
        \draw[dashed, thick](-1.5,-1.5) -- (0,0)  node[midway,above=2pt] {$\alpha$} ;
        \node[isosceles triangle, draw, fill = black, minimum size= 0.15cm, inner sep = 0pt, rotate = 225] at (-0.75,-0.75){} ;
        \filldraw[black](0,0) circle (1.5pt);
        \draw[snake it, thick](0,0) -- (1.5,0) ;
        \filldraw[black](1.5,0) circle (1.5pt);
        \filldraw[black](2.7,-0.8) circle (1.5pt);
        \filldraw[black](2.7,0.8) circle (1.5pt);
        \draw[snake it,thick](2.7,0.8) -- (2.7,-0.8);
        \draw[dashed, thick](2.7,0.8) -- (1.5,0)  node[midway,above=2pt] {$\gamma'$} node[midway]{\slash};
        \node[isosceles triangle, draw, fill = black, minimum size= 0.15cm, inner sep = 0pt, rotate = 225] at (2.1,0.4){} ;
        \draw[dashed, thick](3.5,1.32) -- (2.7,0.8)  node[midway,above=2pt] {$\gamma$} ;
        \node[isosceles triangle, draw, fill = black, minimum size= 0.15cm, inner sep = 0pt, rotate = 225] at (3.1,1.1){} ;
        \draw[-, thick](2.7, -0.8) -- (1.5,0)  node[midway, below=2pt] {$\beta'$} node[midway]{\slash};
        \node[isosceles triangle, draw, fill = black, minimum size= 0.15cm, inner sep = 0pt, rotate = -45] at (2.1,-0.4){} ;
         \draw[-, thick](3.5, -1.32) -- (2.7,-0.8)  node[midway, below=2pt] {$\beta$};
         \node[isosceles triangle, draw, fill = black, minimum size= 0.15cm, inner sep = 0pt, rotate = -40] at (3.1,-1.1){} ;
        \end{tikzpicture}
        \hskip 0.8 cm
        \begin{tikzpicture}[scale=0.55]
        \draw[-, thick](-1.5,1.5) -- (0,0) node[midway,above=2pt] {$\beta$} ; 
        \node[isosceles triangle, draw, fill = black, minimum size= 0.15cm, inner sep = 0pt, rotate = -225] at (-0.75,0.75){}  ;
        \draw[dashed, thick](-1.5,-1.5) -- (0,0)  node[midway,above=2pt] {$\gamma$} ;
        \node[isosceles triangle, draw, fill = black, minimum size= 0.15cm, inner sep = 0pt, rotate = 45] at (-0.75,-0.75){} ;
        \filldraw[black](0,0) circle (1.5pt);
        \draw[snake it, thick](0,0) -- (1.5,0) ; 
        \filldraw[black](1.5,0) circle (1.5pt);
        \filldraw[black](2.7,-0.8) circle (1.5pt);
        \filldraw[black](2.7,0.8) circle (1.5pt);
        \draw[snake it,thick](2.7,0.8) -- (2.7,-0.8);
        \draw[-, thick](3.5,1.32) -- (2.7,0.8)  node[midway,above=2pt] {$\delta$} ;
        \node[isosceles triangle, draw, fill = black, minimum size= 0.15cm, inner sep = 0pt, rotate = 225] at (3.1,1.1){} ;
        \draw[-, thick](2.7,0.8) -- (1.5,0)  node[midway,above=2pt] {$\beta'$} node[midway]{\slash};
        \node[isosceles triangle, draw, fill = black, minimum size= 0.15cm, inner sep = 0pt, rotate = 225] at (2.1,0.4){} ;
        \draw[dashed, thick](2.7, -0.8) -- (1.5,0)  node[midway, below=2pt] {$\gamma'$} node[midway]{\slash};
        \node[isosceles triangle, draw, fill = black, minimum size= 0.15cm, inner sep = 0pt, rotate = -45] at (2.1,-0.4){} ;
         \draw[dashed, thick](3.5, -1.32) -- (2.7,-0.8)  node[midway, below=2pt] {$\alpha$};
         \node[isosceles triangle, draw, fill = black, minimum size= 0.15cm, inner sep = 0pt, rotate = -40] at (3.1,-1.1){} ;
        \end{tikzpicture}
        \\[5pt]
         \begin{tikzpicture}[scale=0.7]
         \node at (-3,0) {(c)};
        \draw[-, thick](-1.75,0.5) -- (-0.75,0.5) node[midway,above=2pt] {$\delta$} ;
        \draw[-, thick](-0.75,0.5) -- (0.75,0.5) node[midway,above=2pt] {$\beta'$} node[midway]{\slash} ;
        \draw[-, thick](1.75,0.5) -- (0.75,0.5) node[midway,above=2pt] {$\alpha$} ;

        \draw[dashed, thick](-1.75,-0.5) -- (-0.75,-0.5) node[midway,above=2pt] {$\beta$} ;
        \draw[dashed, thick](-0.75,-0.5) -- (0.75,-0.5) node[midway,above=2pt] {$\gamma'$} node[midway]{\slash};
        \draw[dashed, thick](1.75,-0.5) -- (0.75,-0.5) node[midway,above=2pt] {$\gamma$};

        \draw[snake it, thick](-0.75,-0.5) -- (-0.75,0.5) node[midway, left=2pt]{};
        \draw[snake it, thick](0.75,-0.5) -- (0.75,0.5) node[midway, right=2pt]{};

        \filldraw[black](-0.75,-0.5) circle (1.5pt);
        \filldraw[black](0.75,-0.5) circle (1.5pt);
        \filldraw[black](-0.75,0.5) circle (1.5pt);
        \filldraw[black](0.75,0.5) circle (1.5pt);

        \node[isosceles triangle, draw, fill = black, minimum size= 0.15cm, inner sep = 0pt, rotate=0] at (0,0.5){} ;
        \node[isosceles triangle, draw, fill = black, minimum size= 0.15cm, inner sep = 0pt, rotate=0] at (-1.25,0.5){} ;
        \node[isosceles triangle, draw, fill = black, minimum size= 0.15cm, inner sep = 0pt, rotate=0] at (1.25,0.5){} ;
        
        \node[isosceles triangle, draw, fill = black, minimum size= 0.15cm, inner sep = 0pt, rotate=180] at (0,-0.5){} ;
        \node[isosceles triangle, draw, fill = black, minimum size= 0.15cm, inner sep = 0pt, rotate=180] at (-1.25,-0.5){} ;
        \node[isosceles triangle, draw, fill = black, minimum size= 0.15cm, inner sep = 0pt, rotate=180] at (1.25,-0.5){} ;
    \end{tikzpicture}
    \\[8pt]
    \begin{tikzpicture}[scale=0.7]
     \node at (-3,0) {(d)};
        \draw[-, thick](-1.75,0.5) -- (-0.75,0.5) node[midway,above=2pt] {$\delta$} ;
        \draw[-, thick](-0.75,0.5) -- (0.75,0.5) node[midway,above=2pt] {$\beta'$}node[midway]{\slash};
        \draw[-, thick](1.75,0.5) -- (0.75,0.5)node[midway,above=2pt] {$\alpha$} ;

        \draw[dashed, thick](-1.75,-0.5) -- (-0.75,-0.5)node[midway,above=2pt] {$\gamma$} ;
        \draw[dashed, thick](-0.75,-0.5) -- (0.75,-0.5) node[midway,above=2pt] {$\gamma'$} node[midway]{\slash} ;
        \draw[dashed, thick](1.75,-0.5) -- (0.75,-0.5)node[midway,above=2pt] {$\beta$} ;

        \draw[snake it, thick](-0.75,-0.5) -- (-0.75,0.5) node[midway, left=2pt]{};
        \draw[snake it, thick](0.75,-0.5) -- (0.75,0.5) node[midway, right=2pt]{};

        \filldraw[black](-0.75,-0.5) circle (1.5pt);
        \filldraw[black](0.75,-0.5) circle (1.5pt);
        \filldraw[black](-0.75,0.5) circle (1.5pt);
        \filldraw[black](0.75,0.5) circle (1.5pt);

        \node[isosceles triangle, draw, fill = black, minimum size= 0.15cm, inner sep = 0pt, rotate=0] at (0,0.5){} ;
        \node[isosceles triangle, draw, fill = black, minimum size= 0.15cm, inner sep = 0pt, rotate=0] at (-1.25,0.5){} ;
        \node[isosceles triangle, draw, fill = black, minimum size= 0.15cm, inner sep = 0pt, rotate=0] at (1.25,0.5){} ;
        
        \node[isosceles triangle, draw, fill = black, minimum size= 0.15cm, inner sep = 0pt, rotate=0] at (0,-0.5){} ;
        \node[isosceles triangle, draw, fill = black, minimum size= 0.15cm, inner sep = 0pt, rotate=0] at (-1.25,-0.5){} ;
        \node[isosceles triangle, draw, fill = black, minimum size= 0.15cm, inner sep = 0pt, rotate=0] at (1.25,-0.5){} ;
    \end{tikzpicture}
    \hskip 0.25cm
    \begin{tikzpicture}[scale=0.7]
   
        \draw[-, thick](-1.75,0.5) -- (-0.75,0.5) node[midway,above=2pt] {$\delta$};
        \draw[dashed, thick](-0.75,0.5) -- (0.75,0.5) node[midway,above=2pt] {$\gamma'$} node[midway]{\slash};
        \draw[-, thick](1.75,0.5) -- (0.75,0.5)node[midway,above=2pt] {$\beta$} ;

        \draw[dashed, thick](-1.75,-0.5) -- (-0.75,-0.5)node[midway,above=2pt] {$\gamma$} ;
        \draw[-, thick](-0.75,-0.5) -- (0.75,-0.5) node[midway,above=2pt] {$\beta'$}node[midway]{\slash} ;
        \draw[dashed, thick](1.75,-0.5) -- (0.75,-0.5) node[midway,above=2pt] {$\alpha$};

        \draw[snake it, thick](-0.75,-0.5) -- (-0.75,0.5) node[midway, left=2pt]{};
        \draw[snake it, thick](0.75,-0.5) -- (0.75,0.5) node[midway, right=2pt]{};

        \filldraw[black](-0.75,-0.5) circle (1.5pt);
        \filldraw[black](0.75,-0.5) circle (1.5pt);
        \filldraw[black](-0.75,0.5) circle (1.5pt);
        \filldraw[black](0.75,0.5) circle (1.5pt);

        \node[isosceles triangle, draw, fill = black, minimum size= 0.15cm, inner sep = 0pt, rotate=0] at (0,0.5){} ;
        \node[isosceles triangle, draw, fill = black, minimum size= 0.15cm, inner sep = 0pt, rotate=0] at (-1.25,0.5){} ;
        \node[isosceles triangle, draw, fill = black, minimum size= 0.15cm, inner sep = 0pt, rotate=0] at (1.25,0.5){} ;
        
        \node[isosceles triangle, draw, fill = black, minimum size= 0.15cm, inner sep = 0pt, rotate=0] at (0,-0.5){} ;
        \node[isosceles triangle, draw, fill = black, minimum size= 0.15cm, inner sep = 0pt, rotate=0] at (-1.25,-0.5){} ;
        \node[isosceles triangle, draw, fill = black, minimum size= 0.15cm, inner sep = 0pt, rotate=0] at (1.25,-0.5){} ;
    \end{tikzpicture}
    \\ [8pt]
    \begin{tikzpicture}[scale=0.7]
     \node at (-3,0) {(e)};
        \draw[-, thick](-1.75,0.5) -- (-0.75,0.5) node[midway,above=2pt] {$\delta$};
        \draw[-, thick](-0.75,0.5) -- (0.75,0.5) node[midway,above=2pt] {$\beta'$} node[midway]{\slash};
        \draw[dashed, thick](1.75,0.5) -- (0.75,0.5) node[midway,above=2pt] {$\alpha$};

        \draw[dashed, thick](-1.75,-0.5) -- (-0.75,-0.5)node[midway,above=2pt] {$\gamma$} ;
        \draw[dashed, thick](-0.75,-0.5) -- (0.75,-0.5)node[midway,above=2pt] {$\gamma'$}node[midway]{\slash}  ;
        \draw[-, thick](1.75,-0.5) -- (0.75,-0.5)node[midway,above=2pt] {$\beta$} ;

        \draw[snake it, thick](-0.75,-0.5) -- (-0.75,0.5) node[midway, left=2pt]{};
        \draw[snake it, thick](0.75,-0.5) -- (0.75,0.5) node[midway, right=2pt]{};

        \filldraw[black](-0.75,-0.5) circle (1.5pt);
        \filldraw[black](0.75,-0.5) circle (1.5pt);
        \filldraw[black](-0.75,0.5) circle (1.5pt);
        \filldraw[black](0.75,0.5) circle (1.5pt);

        \node[isosceles triangle, draw, fill = black, minimum size= 0.15cm, inner sep = 0pt, rotate=0] at (0,0.5){} ;
        \node[isosceles triangle, draw, fill = black, minimum size= 0.15cm, inner sep = 0pt, rotate=0] at (-1.25,0.5){} ;
        \node[isosceles triangle, draw, fill = black, minimum size= 0.15cm, inner sep = 0pt, rotate=0] at (1.25,0.5){} ;
        
        \node[isosceles triangle, draw, fill = black, minimum size= 0.15cm, inner sep = 0pt, rotate=0] at (0,-0.5){} ;
        \node[isosceles triangle, draw, fill = black, minimum size= 0.15cm, inner sep = 0pt, rotate=0] at (-1.25,-0.5){} ;
        \node[isosceles triangle, draw, fill = black, minimum size= 0.15cm, inner sep = 0pt, rotate=0] at (1.25,-0.5){} ;
    \end{tikzpicture}
    \hskip 0.25 cm
    \begin{tikzpicture}[scale=0.7]
        \draw[-, thick](-1.75,0.5) -- (-0.75,0.5)node[midway,above=2pt] {$\delta$} ;
        \draw[dashed, thick](-0.75,0.5) -- (0.75,0.5)node[midway,above=2pt] {$\gamma'$} node[midway]{\slash};
        \draw[dashed, thick](1.75,0.5) -- (0.75,0.5) node[midway,above=2pt] {$\alpha$} ;

        \draw[dashed, thick](-1.75,-0.5) -- (-0.75,-0.5) node[midway,above=2pt] {$\gamma$};
        \draw[-, thick](-0.75,-0.5) -- (0.75,-0.5) node[midway,above=2pt] {$\beta'$} node[midway]{\slash};
        \draw[-, thick](1.75,-0.5) -- (0.75,-0.5) node[midway,above=2pt] {$\beta$};

        \draw[snake it, thick](-0.75,-0.5) -- (-0.75,0.5) node[midway, left=2pt]{};
        \draw[snake it, thick](0.75,-0.5) -- (0.75,0.5) node[midway, right=2pt]{};

        \filldraw[black](-0.75,-0.5) circle (1.5pt);
        \filldraw[black](0.75,-0.5) circle (1.5pt);
        \filldraw[black](-0.75,0.5) circle (1.5pt);
        \filldraw[black](0.75,0.5) circle (1.5pt);

        \node[isosceles triangle, draw, fill = black, minimum size= 0.15cm, inner sep = 0pt, rotate=0] at (0,0.5){} ;
        \node[isosceles triangle, draw, fill = black, minimum size= 0.15cm, inner sep = 0pt, rotate=0] at (-1.25,0.5){} ;
        \node[isosceles triangle, draw, fill = black, minimum size= 0.15cm, inner sep = 0pt, rotate=0] at (1.25,0.5){} ;
        
        \node[isosceles triangle, draw, fill = black, minimum size= 0.15cm, inner sep = 0pt, rotate=0] at (0,-0.5){} ;
        \node[isosceles triangle, draw, fill = black, minimum size= 0.15cm, inner sep = 0pt, rotate=0] at (-1.25,-0.5){} ;
        \node[isosceles triangle, draw, fill = black, minimum size= 0.15cm, inner sep = 0pt, rotate=0] at (1.25,-0.5){} ;
    \end{tikzpicture}
    \caption{ Closed loop (a), vertex corrections (b), ladder  of the Peierls scattering channel (c) and of the electron-electron Cooper scattering channel (d-e).   Slashed fermion lines are in the outer energy shell of the band.\label{loops}}
    \end{figure}
    
In first order for electron-hole scattering at momentum transfer ${\bm q}_{\alpha\gamma}$, also called the Peierls scattering channel, the closed-loop contribution, similar to the diagram of Fig.~\ref{loops}-a, does not involve any summation over intermediate phonon frequencies. For perfect nesting conditions, this leads to a logarithmic singularity of the form $\sim g^2 \ln E_F^\alpha/T$, which is cut off by the high energy $E_F^\alpha =v_F^\alpha k_F^\alpha$, which can be identified with half of the bandwidth. In contrast, perturbative terms of the same Peierls channel, but coming from vertex corrections such as those in Fig.~\ref{loops}-b or ladder terms of Fig.~\ref{loops}-c involve intermediate phonon frequency summations, a situation also found in the case of ladder diagrams of the electron-electron or Cooper scattering channel, as shown in Fig.~\ref{loops}-d,e. All of these lead to corrections with a singularity $\sim g^2 \ln \omega_D/T$ with  weaker energy cutoff $\omega_D$,  a scale that marks the onset of corrections from quantum phonons. 

We shall incorporate these two-cutoff features into the Wilsonian perturbative RG approach \cite{Bourbon91,Bourbon03}. This procedure, which will be carried out at an arbitrary temperature $T$, can be summarized as the iterative partial trace integration for the partition function of electron states in an infinitesimal energy shell $E_0(\ell)d\ell/2 $ on either side of the Fermi surface, where $E_0(\ell)= 2E_F e^{-\ell}$ is the bandwidth in step $\ell$ and is assumed to be the same for each parallel band. The partial trace operation for the partition function is written as follows
\begin{align}
    Z[h] \sim & \! \iint [{\cal D}\psi^*{\cal D}\psi]_<    \, e^{S_\ell}\!  \iint [{\cal D}\psi^*{\cal D}\psi]_{d\ell}\,  e^{S_{0,d\ell}} \, e^{S_{I,d\ell} + S_{h,d\ell}} \cr
   \propto &\iint [{\cal D}\psi^*{\cal D}\psi]_<   \, e^{S_{\ell+d\ell} }.   
\end{align}
The perturbative corrections from the outer energy shells lead to a recursive relation of the action $S_\ell$ to  $\ell$. At the one-loop level, only the interactions of $S_I$  and the source field $S_h$ are affected, and each of their components will obey a recursion relation. 

Considering first the interaction, we have the following  relation,
\begin{align}
    S_{I,\ell + d\ell} = S_{I,\ell}  + \frac{1}{2}\big\langle S_{I,d\ell}^2\big\rangle\Big|_{\mathrm{D}_{\mathrm{c}}+\mathrm{D}_{\mathrm{v}}+ \mathrm{D}_{\mathrm{l}}},  
\end{align}
where $\langle \dots\rangle $ is a thermodynamic average over  fermion fields, carried out with respect to the free action $S_{0,d\ell}$ of fermions in the outer energy shell. The expression adds the contributions of closed-loop ($\mathrm{D}_{\mathrm{c}}$), vertex ($\mathrm{D}_{\mathrm{v}}$) diagrams of the Peierls channel,
and ladder ($\mathrm{D}_{\mathrm{l}}$) diagrams of both Peierls and Cooper ones; these are displayed in Fig.~\ref{loops}-a, \ref{loops}-b, and \ref{loops}-c,d,e, respectively. Here, since the couplings (\ref{g12}) on the Fermi surface are  independent of the momentum, the loops of each scattering channel and then their interference will be maximized in the RG equations.

The explicit evaluation of the diagrammatic terms in the limit $d\ell\to 0$ leads to the following RG equations for the coupling constants,
\begin{widetext}
    \begin{align}
    \partial_\ell g_1^{\alpha\beta;\gamma\delta}= & \ \sum_{\gamma',\beta'} g_1^{\alpha\beta';\gamma'\delta}g_1^{\gamma'\beta;\gamma\beta'}{I}^{\gamma'\beta'}_{\mathrm{P,c}} & +\sum_{\gamma',\,\beta'}\big(g_1^{\gamma'\beta;\gamma\beta'}g_2^{\beta'\alpha;\gamma'\delta} +g_1^{\alpha\beta';\gamma'\delta}g_2^{\beta\gamma';\gamma\beta'} \big) I_{\mathrm{P,v}}^{\gamma'\beta'}\cr   && \ +\sum_{\gamma'=\beta'}\big(g_2^{\gamma'\beta';\gamma\delta}g_1^{\beta\alpha;\gamma'\beta'}+g_1^{\gamma'\beta';\gamma\delta}g_2^{\beta\alpha;\gamma'\beta'}\big)I_{\mathrm{C}}^{\gamma'\beta'}
    \label{g1flow}\\
      \partial_\ell g_2^{\alpha\beta;\gamma\delta}= & \sum_{\gamma',\beta'}g_2^{\gamma'\beta;\beta'\delta}g_2^{\alpha\beta';\gamma\gamma'}{I}^{\gamma'\beta'}_{\mathrm{P,l}} &+  \sum_{\gamma'=\beta'}\big(g_1^{\gamma'\beta';\gamma\delta}g_1^{\beta\alpha;\gamma'\beta'} + g_2^{\gamma'\beta';\gamma\delta}g_2^{\alpha\beta;\gamma'\beta'}\big)I_{\mathrm{C}}^{\gamma'\beta'}.
      \label{g2flow}
\end{align}
\end{widetext}
 for which the in-band couplings, 
 \begin{equation}
     \frac{g^{\alpha\alpha;\alpha\alpha}_{i=1,2}}{\pi v_F^\alpha} \to g^{\alpha\alpha;\alpha\alpha}_{i=1,2},
 \end{equation}
are normalized by the density of states of the $\alpha$ band, whereas the inter-band  couplings, 
\begin{equation}
     \frac{g^{\alpha\alpha;\beta\beta}_{i=1,2}}{\pi \sqrt{v_F^\alpha v_F^\beta}} \to g^{\alpha\alpha;\beta\beta}_{i=1,2}; \ \  \frac{g^{\alpha\beta;\alpha\beta}_{i=1,2}}{\pi \sqrt{v_F^\alpha v_F^\beta}} \to g^{\alpha\beta;\alpha\beta}_{i=1,2},
 \end{equation}
 are normalized by the geometrical density of states of the $\alpha$ and $\beta$  bands. The  flow equations given above are valid for an arbitrary number of parallel bands.
 
The first contribution to the backscattering flow (\ref{g1flow}) comes from the closed-loop diagram D$_\mathrm{c}$ of the Peierls  channel shown in Fig.~\ref{loops}-a. Evaluated at the static vector $\bar{q}_{\gamma'\beta'}=({\bm q}_{\gamma'\beta'},0)$, it is given by 
the temperature- and antinesting-dependent loop coefficient 
\begin{align}
 {I}_{\mathrm{P,c}}^{\gamma'\beta'} = &  \frac{2T}{N_b}\sqrt{\frac{v_F^{\beta'}}{v_F^{\gamma'}}}\sum_{\omega_{\bar{n}},\bar{k}_b} \partial_\ell\intslashD d\bar{E}_-^{\gamma'} \mathscr{D}^2(0)\cr
 &\hskip 2.5cm \times G^{\beta'}_+(\bar{k} +q_{\gamma'\beta'})G^{\gamma'}_-(\bar{k}) \cr
 = & -\mathfrak{I}_{\mathrm{P}}^{\gamma'\beta'},
 \label{}
\end{align}
where   the barred integral is over the outer energy shell on either side of the Fermi surface. The closed loop diagram has no intermediate phonon frequency summation, so that $\mathscr{D}(0)=1$.  In the intraband situation where $\gamma'=\beta'$, the  coefficient  is  explicitly evaluated in Appendix A and is given by  
\begin{align}
&\mathfrak{I}_{\mathrm{P}}^{\gamma'\gamma'} = \cr
     \frac{1}{2} & \Big\langle \frac{\tanh [e^{-\ell}/2\bar{T}] + \tanh [e^{-\ell}/2\bar{T} - 2({\bar{t}'_b}\!^{\gamma'}/\bar{T})\cos 2\bar{k}_b]}{  1 + 2{\bar{t}'_b}\!^{\gamma'} e^{\ell} \cos 2\bar{k}_b} \Big\rangle_{\! \bar{k}_b}\cr
     \label{PCoeff}
\end{align}
where $\bar{T}= T/E_F$ and ${\bar{t}'_b}\!^{\gamma'}={\bar{t}'_b}\!^{\gamma'}/E_F $.  When there is no asymmetry between the bands, that is, for $\bar{v}=1$, the expressions for the coefficients  of the intraband  and interband ($\gamma'\ne\beta'$) are identical.  However, when $\bar{v}\ne 1$, the asymmetry introduces additional nesting changes between the bands and the expressions for $\mathfrak{I}_{\mathrm{P}}^{\p\bp}$  and $\mathfrak{I}_{\mathrm{P}}^{\bp\p}$,  given in (\ref{IPloop}) of  Appendix A, then become distinct from the intraband counterpart.

From (\ref{g1flow}), the vertex loops $\mathrm{D}_{\mathrm{v}}$ of the Peierls channel shown in Fig.~\ref{loops}-b bring corrections to the backscattering amplitude. Evaluated at $\bar{q}_{\gamma'\beta'}=({\bm q}_{\gamma'\beta'},0)$, it takes the form, 
\begin{align}
 {I}^{\gamma'\beta'}_{\mathrm{P,v}} = &-\frac{T}{N_b}\sqrt{\frac{v_F^{\beta'}}{v_F^{\gamma'}}}\sum_{\omega_{\bar{n},\bar{k}_b}} \partial_\ell\intslashD {d\bar{E}^{\gamma'}} \mathscr{D}(\omega_{\bar{n}}-\omega_n)\cr
  & \hskip 2cm \times G^{\beta'}_+(\bar{k}+ \bar{q}_{\gamma'\beta'})G^{\gamma'}_-(\bar{k}) \cr
\simeq & \, \frac{1}{2}X\, \mathfrak{I}_{\mathrm{P}}^{\gamma'\beta'}.
\label{IPv}
\end{align}
The presence of the phonon propagator in the intermediate frequency sum, with $\omega_n$ as an external fermion frequency, indicates that retardation reduces the amplitude of the vertex loop, as already mentioned at the beginning of this section in the context of perturbation theory. Since frequency and electron energy enter on the same footing in the loop, we shall identify the difference in fermion frequencies in the propagator with the electronic energy scale $E_0(\ell)/2$. The expression for $\mathscr{D}$ will thus be replaced by the scaled-dependent cut-off function,
\begin{equation}
   X= (1+ e^{-2\ell}/\bar{\omega}^2_D)^{-1},
\end{equation}
with the same structure as the phonon propagator and where $\bar{\omega}_D= \omega_D/E_F$. This procedure transposes  retardation effects in the static  band energy degrees of freedom; it  reproduces essentially the cutting off of the logarithmic singularity by phonons in perturbation theory.
The last contribution of the Peierls channel to the forward scattering amplitude flow (\ref{g2flow}) comes from the ladder diagram in Fig.~\ref{loops}-c. Its coefficient is given by 
\begin{align}
 {I}^{\gamma'\beta'}_{\mathrm{P,l}} = &-\frac{T}{N_b}\sqrt{\frac{v_F^{\beta'}}{v_F^{\gamma'}}}\sum_{\omega_{\bar{n},\bar{k}_b}} \partial_\ell\intslashD\!\!{d\bar{E}^{\gamma'}} \mathscr{D}(\omega_{\bar{n}}-\omega_n)  \mathscr{D}(\omega_{\bar{n}}-\omega_{n'})\cr
  & \hskip 2cm \times G^{\beta'}_+(\bar{k}+ \bar{q}_{\gamma'\beta'})G^{\gamma'}_-(\bar{k}) \cr
\simeq & \frac{1}{2}X^2\, \mathfrak{I}_{\mathrm{P}}^{\gamma'\beta'}.
\label{Ipl}
\end{align}
This ladder loop comprises the product of two phonon propagators, with $\omega_n $ and $\omega_{n'} $ as external fermion frequencies,  and will  then be  replaced by the square of the cutoff function.

The last logarithmic contributions come from the ladder terms of Fig.~\ref{loops}-d,e for the Cooper scattering channel and contribute to the flow of the forward and backscattering amplitudes, respectively. The related coefficient, when evaluated at zero pair momentum and frequency, only concerns intraband contributions for $\gamma'=\beta'$ and takes the form 
\begin{align}
 {I}^{\gamma'\gamma'}_{\mathrm{C,l}} = &-\frac{T}{N_b}\sqrt{\frac{v_F^{\beta'}}{v_F^{\gamma'}}}\sum_{\omega_{\bar{n}},\bar{k}_b} \!\partial_\ell\intslashD \! {d\bar{E}^{\gamma'}} \mathscr{D}^2(\omega_{\bar{n}}-\omega_{n,n'}) \cr 
& \hskip 2 cm \times G_+^{\gamma'}(\bar{k})G_-^{\gamma'}(-{\bar k}) \cr
 \simeq & - \frac{1}{2}X^2 \, \mathfrak{I}_{\mathrm{C}}^{\gamma'\gamma'},
 \label{Icl}
\end{align}
which, as a ladder term,  becomes also dependent on the square of the cutoff function $X$.  The Cooper loop coefficient is 
\begin{equation}
    \mathfrak{I}_{\mathrm{C}}^{\gamma'\gamma'} = \tanh[e^{-\ell}/2\bar{T}],
    \label{Ic1b}
\end{equation}
which is only affected by temperature. 
\begin{figure}
    \centering
    \begin{tikzpicture}[scale=0.5]
     \node at (-2.4,0) {(a)};
    \draw[dotted, thick](-1.5,0) -- (0,0) node[midway,above=2pt] {$h$} ;
    \draw[dashed, thick](2.2,0) arc (0:-180:0.95) node[midway,below=2pt] {$\alpha$} node[midway]{\slash};
    \draw[-, thick](2.2,0) arc (0:180:0.95) node[midway,above=2pt] {$\delta$} node[midway]{\slash};
    \node[isosceles triangle, draw, fill = black, minimum size= 0.12cm, inner sep = 0pt, rotate = 0] at (1.22,0.95){} ;
    \node[isosceles triangle, draw, fill = black, minimum size= 0.12cm, inner sep = 0pt, rotate = 180] at (1.22,-0.95){} ;
    \draw[snake it,thick](2.2,0) -- (3.5,0);
    \node[isosceles triangle, draw, fill = gray, minimum size= 0.24cm, inner sep = 0pt, rotate = -180] at (0.2,0){}  ;
    \draw[dashed, thick](5.5,0.8) -- (3.5,0)  node[midway,above=2pt] {$\gamma$} ;
    \node[isosceles triangle, draw, fill = black, minimum size= 0.13cm, inner sep = 0pt, rotate = 216] at (4.6,0.45){} ;
    \node[isosceles triangle, draw, fill = black, minimum size= 0.13cm, inner sep = 0pt, rotate = -33] at (4.6,-0.43){} ;
     \draw[-, thick](5.5,-0.8) -- (3.5,0)  node[midway,below=2pt] {$\beta$} ;
     \filldraw[black](3.5,0) circle (1.5pt);
    \end{tikzpicture}
    \hskip 1.0cm
    \begin{tikzpicture}[scale=0.6]
    \draw[dotted, thick](-1.5,0) -- (0,0) node[midway,above=2pt] {$h$} ;
    \node[isosceles triangle, draw, fill = gray, minimum size= 0.24cm, inner sep = 0pt, rotate = -180] at (0.2,0){}  ;
    \draw[-, thick](1.5,0.8) -- (0.23,0.15)  node[midway,above=2pt] {$\delta$} node[midway]{\slash};
    \node[isosceles triangle, draw, fill = black, minimum size= 0.13cm, inner sep = 0pt, rotate = 36] at (0.75,0.4){} ;
    \draw[dashed, thick](1.5,-0.8) -- (0.23,-0.15)  node[midway,below=2pt] {$\alpha$} node[midway]{\slash};
    \node[isosceles triangle, draw, fill = black, minimum size= 0.13cm, inner sep = 0pt, rotate = -213] at (0.75, -0.4){} ;
    \draw[snake it,thick](1.5,-0.8) -- (1.5,0.8);
    \draw[-, thick](2.5,1.25) -- (1.52,0.8)  node[midway,above=2pt] {$\beta$};
    \draw[dashed, thick](2.5,-1.2) -- (1.52,-0.8)  node[midway,below=2pt] {$\gamma$};
    \node[isosceles triangle, draw, fill = black, minimum size= 0.13cm, inner sep = 0pt, rotate = -213] at (1.99,-.99){} ;
    \node[isosceles triangle, draw, fill = black, minimum size= 0.13cm, inner sep = 0pt, rotate = 33] at (1.99, 1.0){}; 
    \end{tikzpicture}
    \\ [8pt]
    \begin{tikzpicture}[scale=0.6]
    \node at (-2.4,0) {(b)};
    \draw[dotted, thick](-1.5,0) -- (0,0) node[midway,above=2pt] {$h$} ;
    \node[isosceles triangle, draw, fill = gray, minimum size= 0.24cm, inner sep = 0pt, rotate = -180] at (0.2,0){}  ;;
    \draw[dashed, thick](1.5,0.8) -- (0.23,0.15)  node[midway,above=2pt] {$\beta$} node[midway]{\slash};
    \node[isosceles triangle, draw, fill = black, minimum size= 0.13cm, inner sep = 0pt, rotate = 36] at (0.75,0.4){} ;
    \draw[-, thick](1.5,-0.8) -- (0.23,-0.15)  node[midway,below=2pt] {$\beta$} node[midway]{\slash};
    \node[isosceles triangle, draw, fill = black, minimum size= 0.13cm, inner sep = 0pt, rotate = -33] at (0.75, -0.4){} ;
    \draw[snake it,thick](1.5,-0.8) -- (1.5,0.8);
    \draw[-, thick](2.5,1.25) -- (1.55,0.8)  node[midway,above=2pt] {$\alpha$};
    \draw[dashed, thick](2.5,-1.2) -- (1.52,-0.8)  node[midway,below=2pt] {$\alpha$};
    \node[isosceles triangle, draw, fill = black, minimum size= 0.13cm, inner sep = 0pt, rotate = -33] at (1.99,-.99){} ;
    \node[isosceles triangle, draw, fill = black, minimum size= 0.13cm, inner sep = 0pt, rotate = 35] at (1.99, 1.0){}; 
    \end{tikzpicture}
    \hskip 1.95cm
    \begin{tikzpicture}[scale=0.6]
    \draw[dotted, thick](-1.5,0) -- (0,0) node[midway,above=2pt] {$h$} ;
    \node[isosceles triangle, draw, fill = gray, minimum size= 0.24cm, inner sep = 0pt, rotate = -180] at (0.2,0){}  ;;
    \draw[dashed, thick](1.5,0.8) -- (0.23,0.15)  node[midway,above=2pt] {$\beta$} node[midway]{\slash};
    \node[isosceles triangle, draw, fill = black, minimum size= 0.13cm, inner sep = 0pt, rotate = 36] at (0.75,0.4){} ;
    \draw[-, thick](1.5,-0.8) -- (0.23,-0.15)  node[midway,below=2pt] {$\beta$} node[midway]{\slash};
    \node[isosceles triangle, draw, fill = black, minimum size= 0.13cm, inner sep = 0pt, rotate = -33] at (0.75, -0.4){} ;
    \draw[snake it,thick](1.5,-0.8) -- (1.5,0.8);
    \draw[dashed, thick](2.5,1.2) -- (1.52,0.8)  node[midway,above=2pt] {$\alpha$};
    \draw[-, thick](2.5,-1.25) -- (1.52,-0.8)  node[midway,below=2pt] {$\alpha$};
    \node[isosceles triangle, draw, fill = black, minimum size= 0.13cm, inner sep = 0pt, rotate = -33] at (1.99,-1.02){} ;
    \node[isosceles triangle, draw, fill = black, minimum size= 0.13cm, inner sep = 0pt, rotate = 36] at (1.99, 0.99){}; 
    \end{tikzpicture}
    
    \caption{One-loop outer energy shell corrections to source field pair vertex parts (gray triangles) of  the susceptibility, (a): $z_{\mathrm{CDW}}^{\alpha\beta}$ for CDW,   (b):  $z_{\mathrm{SC
    }}$ for SC.} 
    \label{zh}
    \end{figure}
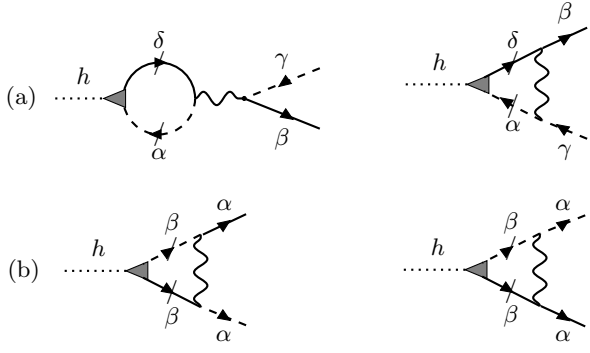

We close this section by considering the flow of the source field terms of $S_h$. At the one-loop level, we have the recursion relation, 
\begin{align}
    S_{h,\ell +d\ell} = S_{h,\ell} + \langle S_{I,d\ell}S_{h,d\ell}\rangle\Big|_{\mathrm{D_{h,c'}}+\mathrm{D_{h,v'}}+\mathrm{D_{h,l'}}},
\end{align}
where are the first two diagrams $\mathrm{D_{h,c'}} $ and  $\mathrm{D_{h,v'}} $ of  the outer energy shell  depicted in Fig.~\ref{zh}-a. These contribute to the pair-vertex coefficient $z^{\gamma\beta}_{\mathrm{CDW}}$ entering  the temperature-dependent susceptibility of CDW,  
\begin{equation}
    \chi_{\mathrm{CDW}}({\bm q}_{\gamma\beta}) = \int_0^\infty \big(z^{\gamma\beta}_{\mathrm{CDW}}\big)^2 \, \mathfrak{I}_P^{\gamma\beta}d\ell,
    \label{KiCDW}
\end{equation}
 evaluated at $\bar{q}_{\gamma\beta}=({\bm q}_{\gamma\beta},0)$. The diagrams with a phonon line are related  to the closed-loop and vertex and lead to the equation
\begin{align}
    \partial_\ell z^{\gamma\beta}_{\mathrm{CDW}} =  z^{\alpha\delta}_{\mathrm{CDW}}\big(g_1^{\alpha\beta;\gamma\delta} I_{\mathrm{P,c'}}^{\alpha\delta} + g_2^{\beta\alpha;\gamma\delta} I^{\alpha\delta}_{\mathrm{P,v'}}\big),
    \label{zcdw}
\end{align}
in which we have the expressions $I^{\alpha\delta}_{\mathrm{P,c'}}= -\mathfrak{I}_P^{\alpha\delta}$  and $I^{\alpha\delta}_{\mathrm{P,v'}}=X\mathfrak{I}_P^{\alpha\delta}/2$. Finally, the last correction to $S_h$ comes from the ladder diagram of Fig.~\ref{zh}-b for the coefficient $z_{\mathrm{SC}}^{\alpha\alpha}$ of superconducting static susceptibility 
\begin{equation}
  \chi^{\alpha\alpha}_{\mathrm{SC}} = \int_0^\infty \big(z_{\mathrm{SC}}^{\alpha\alpha}\big)^2 \mathfrak{I}_C^{\alpha\alpha}\,d\ell,
\label{KiSC}  
\end{equation} 
which can only be intraband when evaluated at zero pair momentum. The ladder diagram of Fig.~\ref{zh}-b,  with  one phonon exchange line, yields the flow equation of $z_{\mathrm{SC}}^{\alpha\alpha}$,
\begin{align}
    \partial _\ell \ln z_{\mathrm{SC}}^{\alpha\alpha} = & \   (g_1^{\alpha\alpha;\alpha\alpha} + g_2^{\alpha\alpha;\alpha\alpha})I_{\mathrm{C,l'}}^{\alpha\alpha}\cr
     &  +\ (g_1^{\alpha\alpha;\beta\beta} + g_2^{\alpha\alpha;\beta\beta})I_{\mathrm{C,l'}}^{\beta\beta},
     \label{zsc}
\end{align}
where the expression for $I_{\mathrm{C,l'}}^{\alpha\alpha}= -X\mathfrak{I_{\mathrm{C}}}^{\alpha\alpha}/2$ is linear in the cutoff function.  Here, both intraband and interband $(\alpha\ne \beta)$ couplings contribute to intraband superconductivity. 
\section{Results}
\label{Results}
\subsection{Separate bands: adiabatic   limit}
\label{SecAdia1b}

We first consider the situation where the bands are separated. In the adiabatic limit where $\omega_D\to0$, the phonons are classical and the characteristic step scale $\ell_D=\ln E_F/\omega_D\to \infty $ at which the phonon exchange becomes quantum mechanically coherent and $X$ becomes appreciable goes to infinity.  The flow equations of each band at $X\to 0$ then contain only the closed loop diagram. For the  amplitude of backscattering  $g_1^{\alpha\alpha;\alpha\alpha}$ at momentum transfer ${\bm q}_\alpha=(2k_F^\alpha,\pi)$, written as $g_1^\alpha$ in this subsection, the flow equation (\ref{g1flow}) reduces to 
\begin{align}
    \partial_\ell g_1^{\alpha}=  \ -(g_1^{\alpha})^2 \mathfrak{I}^\alpha_P,
    \label{g1adia_1b}
\end{align}
where $\mathfrak{I}^\alpha_P$ is the coefficient of the Peierls loop given in (\ref{PCoeff}) and $\mathfrak{g}_1^\alpha \equiv {g_1^{\alpha}(\ell=0)  <0}$ as the initial attractive value. 

Regarding the response functions, only the CDW susceptibility evaluated at the wave vector ${\bm q}_\alpha$ of charge modulation can be considered for each band $\alpha=\p,\bp$. It is the only one that can develop a singularity in the adiabatic $X\to 0$ limit. Corrections to the pair-vertex function $z_{\rm CDW}^\alpha$ at ${\bm q}_\alpha$ come exclusively from the closed-loop diagram shown in Fig.~\ref{zh}-(a) for each band. Thus, at the one-loop level from (\ref{zcdw}), one has  
\begin{equation}
    \partial_\ell\ln z^\alpha_{\rm CDW} = - g_1^{\alpha}\mathfrak{I}^\alpha_P, 
    \label{zcdwadia}
\end{equation}
with the initial condition $z^\alpha_{\rm CDW}=1$ at $\ell=0$.  The solution of (\ref{zcdwadia}) and (\ref{g1adia_1b}) as a function of temperature can be obtained when $\ell\to \infty$: 
\begin{equation}
  g_1^{\alpha}= \frac{\mathfrak{g}_1^{\alpha}}{1+ \mathfrak{g}_1^{\alpha} \chi_0^\alpha} 
  \label{g1badia}
\end{equation}
and
  \begin{equation}
    z^\alpha_{\rm CDW}  
  = \frac{1}{1+ \mathfrak{g}_1^{\alpha}\chi_0^\alpha}.
   \label{zgadia1b}
  \end{equation}
Here, the temperature- and antinesting-dependent bare normalized electron susceptibility $\chi_0^\alpha$ at ${\bm q}_\alpha$ can be analytically evaluated,
\begin{align}
  &\chi_0^\alpha =   \int_0^\infty \mathfrak{I}_P^\alpha d\ell= \ln(1.14E_F/T) \cr
     & + \psi ({1}/{2})  -\frac{1}{2} \big[\big\langle \psi\big(1/2 + i(\pi T)^{-1} {t}_b'^{\alpha}\cos 2\bar{k}_b\big)\big\rangle_{\bar{k}_b} + \mathrm{c.c}\big],\cr
     \label{ki0a}
\end{align}
where $\psi(x)$ as the digamma function.  

The above expressions (\ref{g1badia}-\ref{zgadia1b}) show the expected simple pole structure of the mean-field  approach to the instability of the electron system against the CDW$_\alpha$ ordering at ${\bm q}_\alpha$. 
\begin{figure}[b]
\includegraphics[width=\linewidth]{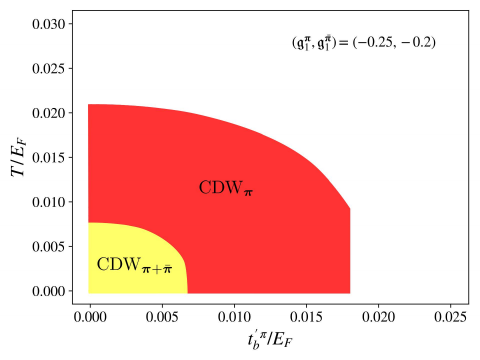} 
\caption{\label{fig_1badia} Phase diagram of  model with separate bands in the adiabatic zero phonon frequency  limit.}
\end{figure}
The search of the characteristic ordering temperature $T_{\alpha}$ at which a singularity occurs simultaneously in these expressions  leads to the phase diagram in Fig.~\ref{fig_1badia}. In perfect nesting conditions, when ${t}_b'^{\alpha}\to 0$, $T_{\alpha}$ takes its maximum value ${T^0_{\alpha} = 1.14 E_F \exp[-1/| \mathfrak{g}_1^\alpha|]}$ . By raising the nesting mismatch ${t}_b'^{\alpha}$  for each band, the Peierls loop coefficient and, with it, $T_{\alpha}$, are monotonically decreasing until a threshold ${t}_{b,c}'^{\alpha}$ is approached from below where $T_{\alpha}$  is critically driven to zero, as shown in Fig.~\ref{fig_1badia} (See the comment \footnote{Here we ignore the possibility of a CDW instability occurring at a  modulation wave vector different than ${\bm q}_\alpha$ at very low temperature and over a small $t'_b$ interval above the threshold ${t}_{b,c}'^{\alpha}$. See for instance, Y. Hasegawa and H. Fukuyama, J. Phys. Soc. Jpn. {\bf 55}, 3978 (1986) and G. Montambaux, Phys. Rev. {\bf 38}, 4788 (1988). It is worth noting that beyond the adiabatic limit when phonon frequency is finite, this additional density-wave state would be suppressed by the superconductivity dome.    }). These one-band results are congruent with those derived many years ago in the framework of single band adiabatic mean-field theory of weakly coupled Peierls chains \cite{Horovitz75}. The monotonous decrease followed by the critical vanishing of $T_\alpha$ simulates the influence of applied pressure on the Fermi surface in concrete situations \cite{Monceau12,Pouget24}.      
\subsection{Separate bands: finite phonon frequency}

We now consider the case where for separate bands the phonon frequency $\omega_D$ is finite but smaller than the electronic scale $E_F$. For each band, this leads to a finite RG step $\ell_D$, at which the cutoff function $X$  becomes finite as $\ell$ approaches $\ell_D$ and beyond. One-loop vertex and ladder corrections with intermediate phonon summations then start to contribute as antiadiabatic corrections to the RG flow. For each band $\alpha$, the flow equations become
\begin{align}
\label{1bg1nadia}
    &\partial_\ell g_1^{\alpha}=   - (g_1^{\alpha})^2 \mathfrak{I}^\alpha_P  + X g_1^{\alpha}g_2^{\alpha}\mathfrak{I}^\alpha_P  - X^2 g_1^{\alpha}g_2^{\alpha} \mathfrak{I}_C^\alpha \\ \cr
 \text{and} \cr
& \partial_\ell g_2^{\alpha}=   \ \frac{1}{2} X^2\big[(g_2^{\alpha})^2\mathfrak{I}^\alpha_P  
-\big( (g_1^{\alpha})^2 + (g_2^{\alpha})^2 \big)\mathfrak{I}_C^\alpha\big],\cr
\label{1bg2nadia}
\end{align}
for the backward and forward scattering amplitudes within the bands that are also simply rewritten as $g_{1,2}^{\alpha\alpha;\alpha\alpha}\to g_{1,2}^\alpha$ in this subsection. 

Antiadiabatic corrections also contribute to the pair-vertex part $z_{\rm CDW}^\alpha$ of the ${\bm q}_\alpha $  susceptibility of CDW$_\alpha$ together with emergent Cooper-pairing correlations through the pair-vertex part $z_{\rm SC}^\alpha$ for the susceptibility of SC$_\alpha$ singlet superconductivity  of the band $\alpha$.  From (\ref{zcdw}-\ref{zsc}), one has 
\begin{align}
\label{z1bNadia_CDW}
    \partial_\ell\ln z^\alpha_{\rm CDW} = &\, (-g_1^{\alpha} + X\, g_2^{\alpha}/2) \,\mathfrak{I}^\alpha_P,\\
   \partial_\ell\ln z^\alpha_{\rm SC} = &-\frac{1}{2}X  (g_1^{\alpha} +g_2^{\alpha}) \,\mathfrak{I}^\alpha_C.
    \label{z1bNadia_SC}
\end{align}
According to (\ref{g12}), the initial condition for $g_1^\alpha(\ell=0)\to \mathfrak{g}_1^\alpha $ is finite at a large momentum transfer $\sim {\bm q}_\alpha$, while for  $g_2^\alpha(\ell=0)\to \mathfrak{g}_2^\alpha \sim 0  $, it starts to vanishingly small value at zero momentum transfer but becomes finite at $\ell>0$.
From the integration of (\ref{1bg1nadia}-\ref{1bg2nadia}) and (\ref{z1bNadia_CDW}-\ref{z1bNadia_SC}) up to $\ell \to \infty$, the characteristic ordering temperature $T_{\alpha }$ for CDW$_\alpha$ or $T_{c,\alpha }$ for SC$_\alpha$ associated with the simultaneous singularities  of these equations can be obtained as a function of ${t}_b'^{\alpha}$. 
\begin{figure}[b]
\includegraphics[width=\linewidth]{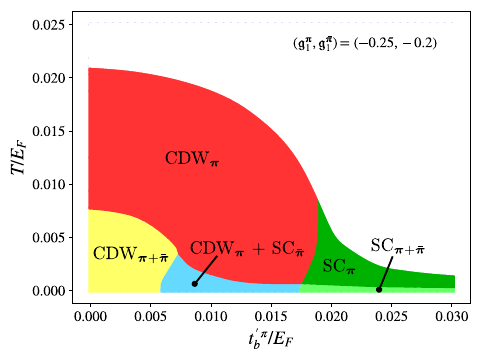} 
\caption{\label{fig_1bNadia} Phase diagram for separated bands at finite phonon frequency, $\bar{\omega}_D=0.033$.}
\end{figure}
The derived phase diagram is shown in Fig.~\ref{fig_1bNadia} for different initial intraband couplings. 

The qualitative new feature compared to the adiabatic results of Fig.~\ref{fig_1badia}  is the presence of superconductivity. It now appears as a new instability of the electron system, emerging from the collapse of the ordered CDW$_\alpha$  state at the critical antinesting value ${t}_{b}'^{*\alpha}$. However, $\mathrm{CDW}_\alpha$ short range quantum correlations remain critically large close to  ${t}_{b}'^{*\alpha}$ where the critical temperature $T_{c,\alpha}$ reaches its maximum value. This indicates that Cooper pairing induced by phonon exchange between electrons is enhanced by the presence of quantum CDW$_\alpha$ correlations. This makes each CDW$_\alpha$ quantum critical point unstable against the formation of a superconducting dome for each band. As antinesting is further enhanced above ${t}_{b}'^{*\alpha}$, these correlations are gradually suppressed and, in turn, are monotonically reducing $T_{c,\alpha}$. Since according to (\ref{Ic1b}), the Cooper loop is not affected by antinesting, $T_{c,\alpha}$, though steadily decreasing, always remains finite even well above ${t}_{b}'^{*\alpha}$, in the end of the superconducting dome for each separate band.  It is worth noting that according to the Figure and as discussed in the Appendix \ref{Boundary},  the  integration  of the RG flow at finite temperature  allows the incursion of superconductivity within the zone of stability of CDW. As the bands do not interact, the ordered  states can  superimpose for  CDW$_{\p +\bp}$ and SC$_{\p +\bp}$ for CDW and SC, respectively, as  shown in Figure~\ref{fig_1bNadia}.

Other features are also worth mentioning. First, at zero ${t}_b'^{\alpha}$ the loop coefficients $\mathfrak{I}_P^\alpha$ and $\mathfrak{I}_C^\alpha $ become equal and lead to important cancelations due to interference in (\ref{1bg1nadia}-\ref{1bg2nadia}).  Since in the example shown, $T_{\alpha }< \omega_D$, the adiabatic character of the CDW$_\alpha$ ordering in (\ref{z1bNadia_CDW}) is altered to some extent by quantum corrections coming from Cooper pair fluctuations. At finite ${t}_b'^{\alpha}$, the amplitude of $\mathfrak{I}_P^\alpha  $ is reduced with respect to $\mathfrak{I}_C^\alpha$. This adds further corrections to the CDW instability at $T_{\alpha}$ due to a change in the interference between density wave and Cooper pairings that favors the latter in (\ref{1bg1nadia}-\ref{1bg2nadia}). At critical $ {t}_{b}'^{*\alpha}\sim T^0_{\alpha }$ where superconductivity emerges on the brink of falling $\mathrm{CDW}_\alpha$  ordering, the critical temperature $T_{c,\alpha}$ is maximum.

\subsection{Two-band: adiabatic limit}
In the two-band situation,  in addition to the backscattering amplitudes within each band, henceforth renamed as $ g_1^\alpha \to g_1^{\alpha\alpha;\alpha\alpha} $, we have the  additional momentum conserving backscattering amplitudes between the bands, $g_1^{\alpha\alpha;\beta\beta} $  and $g_1^{\alpha\beta;\alpha\beta} $ for $\alpha\ne \beta $.

In the adiabatic limit when $\omega_D\to 0$ and $\ell_D \to \infty$, only closed  loops contribute to the flow of each type of coupling.
  \begin{figure*}
\includegraphics[width=1.0\linewidth]{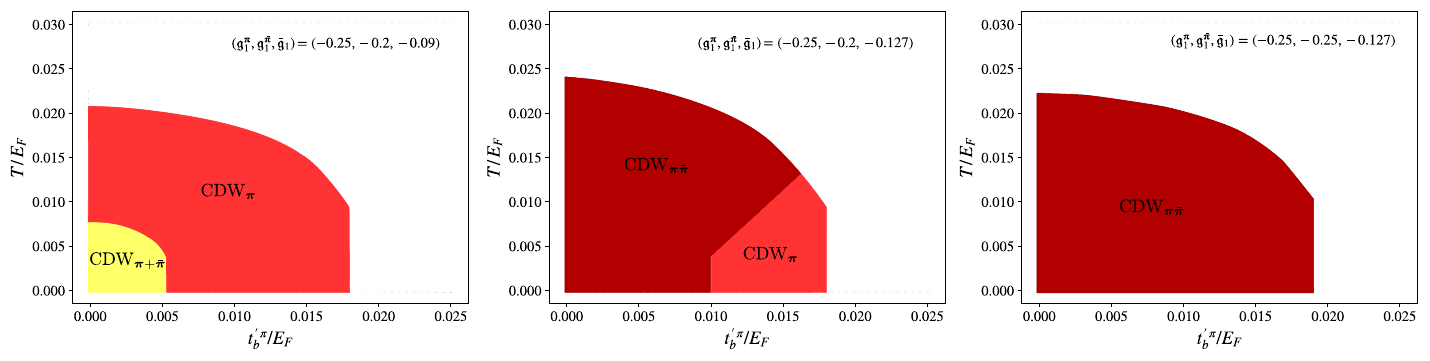}
\caption{Phase diagrams for the coupled  bands in the adiabatic limit of zero phonon frequency ($\bar{\omega}_D=0)$. (a) $  2|\bar{\mathfrak{g}}_1|<  \mathrm{Max}\{|\mathfrak{g}^\alpha_1| \}$ at $\bar{v}=1$  ; (b) $  2|\bar{\mathfrak{g}}_1|\gtrsim   \mathrm{Max}\{|\mathfrak{g}^\alpha_1| \}$ at $\bar{v}=1.25$; (c) $2|\bar{\mathfrak{g}}_1|\gtrsim   \mathrm{Max}\{|\mathfrak{g}^\alpha_1| \}$ at $\bar{v}=1$ . \label{DPhases_Inter_adia}}
\end{figure*}
Within each band, we thus recover the results:
\begin{align}
&\partial_\ell g_1^{\alpha\alpha;\alpha\alpha} =  - (g_1^{\alpha\alpha;\alpha\alpha} )^2\mathfrak{I}_P^{\alpha\alpha} 
\label{intraflowadiapbp}
\end{align}
for the backscattering amplitude at momentum transfer ${\bm q}_{\alpha} =(2k_F^\alpha,\pi) $, and 
\begin{align}
\label{z1Nadia_CDW}
\partial_\ell\ln z^{\alpha\alpha}_{\rm CDW} = &- g_1^{\alpha\alpha;\alpha\alpha}  \,\mathfrak{I}^{\alpha\alpha}_P,
    \end{align}
for the pair-vertex  of the CDW$_\alpha$ susceptibility at the modulation wavevector ${\bm q}_{\alpha} $. These  coincide with  (\ref{g1adia_1b}) 
and (\ref{zcdwadia}) and their respective analytical solutions (\ref{g1badia}) and (\ref{zgadia1b})  for $\alpha=\p $ and $\bp$. Note that, due to the constraint of momentum conservation, no interband coupling contributes to the above equations.

As for the interband scattering amplitudes in the adiabatic limit, their flow equations obtained from (\ref{g1flow}) are  given by the four equations,
\begin{align}
\label{g1Ia}
 &  \!\!\!\!\partial_\ell g_{1}^{\alpha\beta;\alpha\beta}  =   - (g_{1}^{\alpha\beta;\alpha\beta})^2 \mathfrak{I}_P^{\alpha\beta}- g_{1}^{\alpha\alpha;\beta\beta} g_{1}^{\beta\beta;\alpha\alpha} \mathfrak{I}_P^{\beta\alpha},
 \\
 \label{g1Ib}
  & \!\!\!\! \partial_\ell  g_{1}^{\alpha\alpha;\beta\beta}  = -  g_{1}^{\alpha\alpha;\beta\beta} g_{1}^{\beta\alpha;\beta\alpha}  \mathfrak{I}_P^{\beta\alpha} - g_{1}^{\alpha\beta;\alpha\beta}g_{1}^{\alpha\alpha;\beta\beta} \mathfrak{I}_P^{\alpha\beta}  
\end{align}
for $\alpha\ne \beta$ and which involve the  interband loop $\mathfrak{I}_P^{\alpha\beta} $, 
  evaluated at momentum transfer $\bm{q}_I=(k_F^\p+ k_F^{\bp},\pi)$; its   expressions as a function of the antinesting,  temperature and band asymmetry  $\bar{v} $ are given in (\ref{IPloop}) of Appendix A. 
 
 The pair-vertex parts $z^{\alpha\beta}_{\text{CDW}} $ for the  interband susceptibility of  CDW$_{\alpha\beta} $ at the modulation wave vector ${\bm q}_I$ can also be introduced. It follows from the closed loop contribution of (\ref{zcdw}) shown in Fig.~\ref{zh}-(a),    
 \begin{align}
\label{zcdwpbp}
   \partial_\ell  z^{\alpha\beta}_{\text{CDW}} & = - g_1^{\beta\beta;\alpha\alpha} z^{\beta\alpha}_{\text{CDW}}  \mathfrak{I}_P^{\beta\alpha} -g_1^{\alpha\beta;\alpha\beta} z^{\alpha\beta}_{\text{CDW}}  \mathfrak{I}_P^{\alpha\beta}.
   \end{align}
 We note that in the adiabatic limit, the above intraband and interband RG flows are decoupled, which implies that the related possible orderings can occur independently. This is equivalent to a mean-field approach to the instability of the metallic state against the different possibilities of CDW orders. However, in the RG framework, a singularity occurring in the flow signals strong coupling or a gap opening at some energy in one or even two bands whose impact is not included by the present approach   at the one-loop level.  The gap formation will suppress the density of states of the bands involved. These higher order effects can be nevertheless  taken into account  by removing  all related couplings from the flow at lower energy. 

 In the particular case of perfectly parallel bands where $\bar{v}=1$, the flows of inter-band couplings with the same initial conditions, $\bar{\mathfrak{g}}_1\equiv g_1^{\alpha\beta;\alpha\beta}|_{\ell=0}=g_1^{\alpha\alpha;\beta\beta}|_{\ell=0}<0$, together with the pair-vertex terms for the inter-band CDW$_{\alpha\beta}$-susceptibility are, respectively, identical. In this limit, $ \mathfrak{I}_P^{\alpha\beta}=\mathfrak{I}_P^{\alpha\alpha}$, all closed loops become equal and the solutions for (\ref{g1Ia}-\ref{g1Ib}) and (\ref{zcdwpbp})  take the following forms:    
 \begin{align}
 g_1^{\alpha\beta;\alpha\beta} &  = g_1^{\alpha\alpha;\beta\beta} \cr
   &=  \frac{\bar{\mathfrak{g}}_1}{ 1+2\bar{\mathfrak{g}}_1 \chi_0^{\alpha\beta}}
 \end{align}
 and 
 \begin{equation}
     z^{\alpha\beta}_{\text{CDW}}=  \frac{1}{ 1+2\bar{\mathfrak{g}}_1\chi_0^{\alpha\beta}},
 \end{equation}
 where $\chi_0^{\alpha\beta}=\int_0^\infty\mathfrak{I}_P^{\alpha\beta}d\ell$ is the temperature and antinesting dependent interband bare susceptibility evaluated at the wave vector ${\bm q}_I$. Since each interband flow equation above contains twice as many loop contributions compared to the intraband counterparts in (\ref{intraflowadiapbp}-\ref{z1Nadia_CDW}), this doubles the number of couplings involved in the CDW$_{\alpha\beta}$ instability at $\bm{q}_{I}$ that combines both bands. It follows that it is  the ratio $2|\bar{\mathfrak{g}}_1|/\mathrm{Max}\{|\mathfrak{g}^\alpha_1| \}$ that essentially  governs the nature of the phase stabilized.

 For $\bar{v}=1$, the expression for $\chi_0^{\alpha\beta} $ has the same form as given by $\chi_0^\alpha$  in (\ref{ki0a}).
 Thus, when the couplings satisfy the inequality, $  2|\bar{\mathfrak{g}}_1|< \mathrm{Max}\{|\mathfrak{g}^\alpha_1| \} $   the CDW$_{\alpha}$  order ${\bm q}_\alpha$  occurs separately in each band as shown in Fig.~\ref{DPhases_Inter_adia}-(a), a case similar to the isolated band situation discussed in Sect.~\ref{SecAdia1b} and displayed in Fig.~\ref{fig_1badia}. Reversing the  inequality $2 |\bar{\mathfrak{g}}_1|>  \mathrm{Max}\{|\mathfrak{g}^\alpha_1| \}$, it is an inter-band Peierls instability that occurs first. This introduces a strong coupling and the opening of a gap in each band, stopping the flow of intraband couplings and forbidding any intraband  ${\bm q}_\alpha$  ordering to occur. In these conditions and in perfect nesting, for example, one gets $T^0_{\p\bp}= 1.14E_F \exp[-1/(2|\bar{\mathfrak{g}}_1|) ]$ for the maximum critical temperature for interband CDW$_{\p\bp}$  ordering. By increasing the antinesting amplitude, the reduction of $\mathfrak{I}_P^{\alpha\beta}$ leads to the monotonic decay of $T_{\p\bp}$ until its critical drop to zero on approach of a threshold value, as shown in Fig.~\ref{DPhases_Inter_adia}-(c).

When $\bar{v} \ne 1$, there is a departure from band parallelism and an interband electron-hole asymmetry is introduced, which makes, according to (\ref{IPloop}), $\mathfrak{I}_P^{\p\bp}$ and $ \mathfrak{I}_P^{\bp\p}$ differ from $\mathfrak{I}_P^{\alpha\alpha}$.   The temperature scales for possible CDW orders follow from the integration of (\ref{intraflowadiapbp}), (\ref{g1Ia}-\ref{g1Ib}), (\ref{z1Nadia_CDW}) and (\ref{zcdwpbp}) up to $\ell\to\infty$.   In the case where the inequality $ 2|\bar{\mathfrak{g}}_1|> \mathrm{Max}\{|\mathfrak{g}^\alpha_1| \} $ holds for initial conditions, the interchain CDW$_{\p\bp}$  dominates the phase diagram and differs only quantitatively from the previous symmetric situation depicted in Fig.~\ref{DPhases_Inter_adia}-(c) at $\bar{v}=1$. When $ 2|\bar{\mathfrak{g}}_1| \gtrsim \mathrm{Max}\{|\mathfrak{g}^\alpha_1| \} $ and the inequality is close to reverse, the asymmetry  adds nesting deviations between the bands that alter the dominance of the interchain  CDW$_{\p\bp}$ over a part of the phase diagram. This allows the incursion of an intraband  CDW$_\p$ phase, as shown in Fig.~\ref{DPhases_Inter_adia}-(b)   when $\bar{v} >1$  (for $\bar{v}< 1$, $g_1^{\p\p;\p\p}=-0.2 $ and $ g_1^{\bp\bp;\bp\bp}=-0.25$, and the  CDW$_{\bp}$ phase is  inserted instead). As $ 2 |\bar{\mathfrak{g}}_1|$ grows away from $\mathrm{Max}\{|\mathfrak{g}^\alpha_1| \}$ these incursions are gradually suppressed in favor of interband CDW$_{\p\bp}$.

\subsection{Coupled two-band case at finite phonon frequency}
We now examine  the multi-band situation at finite phonon frequency. In the Peierls scattering channel, the central equations (\ref{g1flow}-\ref{g2flow}) yield closed loop intraband and interband  terms of the above adiabatic case that contribute at all $\ell$. To these add the vertex terms of Fig.~\ref{loops}-(b)  which are linear in the cutoff function $X$ and   only emerge from $\ell\sim \ell_D$. As for ladder terms of Fig.~\ref{loops}-(ce) for Peierls and Cooper intra- and inter-band scattering channels which are proportional to $X^2$, their onset in the flow according to (\ref{Ipl}-\ref{Icl}) proceeds almost similarly as a function of  $\ell$.

Thus, collecting all contributions, one can write the flow equations  of intraband couplings 
\begin{widetext}

\begin{align}
    \partial_\ell g_1^{\alpha\alpha;\alpha\alpha} = \, & -(g_1^{\alpha\alpha;\alpha\alpha} )^2\mathfrak{I}_P^{\alpha\alpha} + Xg_1^{\alpha\alpha;\alpha\alpha}g_2^{\alpha\alpha;\alpha\alpha}\mathfrak{I}_P^{\alpha\alpha} - X^2g_1^{\alpha\alpha;\alpha\alpha}g_2^{\alpha\alpha;\alpha\alpha} \mathfrak{I}_C^{\alpha\alpha} - \frac{X^2}{2}\big(g_1^{\beta\beta;\alpha\alpha}g_2^{\alpha\alpha;\beta\beta}+ g_1^{\alpha\alpha;\beta\beta}g_2^{\beta\beta;\alpha\alpha}\big)\mathfrak{I}_C^{\beta\beta}
    \label{g1IntraG}
    \\
    \partial_\ell g_2^{\alpha\alpha;\alpha\alpha} =  &\ \frac{X^2}{2}\Big\{ \big(g_2^{\alpha\alpha;\alpha\alpha}\big)^2 \big(\mathfrak{I}_P^{\alpha\alpha} - \mathfrak{I}_C^{\alpha\alpha}\big) -\big(g_1^{\alpha\alpha;\alpha\alpha}\big)^2 \mathfrak{I}_C^{\alpha\alpha}+ g_2^{\beta\alpha;\beta\alpha}g_2^{\alpha\beta;\alpha\beta}\mathfrak{I}_P^{\beta\beta} - \big(g_2^{\beta\beta;\alpha\alpha}g_2^{\alpha\alpha;\beta\beta}+g_2^{\beta\beta;\alpha\alpha}g_2^{\alpha\alpha;\beta\beta}\big) \mathfrak{I}_C^{\beta\beta}\Big\},
     \label{g2IntraG}
    \end{align}
    where $\alpha=\p,\bp$ and $\beta \ne \alpha$. 
Regarding interband couplings, they  can be grouped together as
\begin{align}
    \partial_\ell g_1^{\alpha\beta;\alpha\beta} = & \, -( g_1^{\alpha\beta;\alpha\beta})^2\mathfrak{I}_P^{\alpha\beta} - g_1^{\alpha\alpha;\beta\beta}g_1^{\beta\beta;\alpha\alpha}\mathfrak{I}_P^{\beta\alpha} + \frac{X}{2}\Big\{\big(g_1^{\beta\beta;\alpha\alpha}g_2^{\alpha\alpha;\beta\beta}+g_1^{\alpha\alpha;\beta\beta}g_2^{\beta\beta;\alpha\alpha}\big)\mathfrak{I}_P^{\beta\alpha} + 2g_1^{\alpha\beta;\alpha\beta}g_2^{\beta\alpha;\alpha\beta}\mathfrak{I}_P^{\alpha\beta}\Big\}\\
   \partial_\ell  g_1^{\alpha\alpha;\beta\beta} =  & \,  -g_1^{\alpha\beta;\alpha\beta}g_1^{\alpha\alpha;\beta\beta} \mathfrak{I}_P^{\alpha\beta} - g_1^{\alpha\alpha;\beta\beta}g_1^{\beta\alpha;\beta\alpha} \mathfrak{I}_P^{\beta\alpha} \cr   & + \frac{X}{2}\Big\{\big(g_1^{\beta\alpha;\beta\alpha}g_2^{\alpha\alpha;\beta\beta}+ g_1^{\alpha\alpha;\beta\beta}g_2^{\alpha\beta;\beta\alpha}\big)\mathfrak{I}_P^{\beta\alpha}  
   +\big(g_1^{\beta\alpha;\beta\alpha}g_2^{\alpha\alpha;\beta\beta}+ g_1^{\alpha\alpha;\beta\beta}g_2^{\alpha\beta;\beta\alpha}\big)\mathfrak{I}_P^{\alpha\beta} \Big\} \cr
   & -\frac{X^2}{2}\Big\{ \big(g_1^{\alpha\alpha;\beta\beta}g_2^{\beta\beta;\beta\beta} + g_1^{\beta\beta;\beta\beta}g_2^{\alpha\alpha;\beta\beta}\big)\mathfrak{I}_C^{\beta\beta}+\big(g_1^{\alpha\alpha;\alpha\alpha}g_2^{\alpha\alpha;\beta\beta} + g_1^{\alpha\alpha;\beta\beta}g_2^{\alpha\alpha;\alpha\alpha}\big)\mathfrak{I}_C^{\alpha\alpha}  \Big\},\cr
\end{align}
for the flow of the four  interband backward scattering amplitudes when $\alpha=\p,\bp$ and $\beta \ne \alpha$, and similarly, 
\begin{align}
    \partial_\ell g_2^{\alpha\beta;\alpha\beta} =  \, \frac{X^2}{2}\Big\{&\big(g_2^{\alpha\beta;\beta\alpha}\big)^2\mathfrak{I}_P^{\beta\alpha} + g_2^{\alpha\alpha;\beta\beta}g_2^{\beta\beta;\alpha\alpha}\mathfrak{I}_P^{\alpha\beta}\Big\}\\
    \partial_\ell g_2^{\alpha\alpha;\beta\beta} =  \, \frac{X^2}{2}\Big\{&g_2^{\beta\alpha;\alpha\beta}g_2^{\alpha\alpha;\beta\beta} \mathfrak{I}_P^{\alpha\beta} + g_2^{\alpha\alpha;\beta\beta} g_2^{\alpha\beta;\beta\alpha}\mathfrak{I}_P^{\beta\alpha}-\big(g_2^{\alpha\alpha;\beta\beta}g_2^{\alpha\alpha;\alpha\alpha} +g_1^{\alpha\alpha;\beta\beta}g_1^{\alpha\alpha;\alpha\alpha}\big)\mathfrak{I}_C^{\alpha\alpha} \cr  &-  \big(g_2^{\alpha\alpha;\beta\beta}g_2^{\beta\beta;\beta\beta} +g_1^{\alpha\alpha;\beta\beta}g_1^{\beta\beta;\beta\beta}\big)\mathfrak{I}_C^{\beta\beta}\Big\},
    \end{align}
    for the flow of the  four forward scattering amplitudes. The initial values noted as  $g_2^{\alpha\beta;\alpha\beta}\big|_{\ell=0} \equiv \bar{\mathfrak{g}}_2\approx  0 $ and $g_2^{\alpha\alpha;\beta\beta}\big|_{\ell=0} \equiv \bar{\mathfrak{g}}'_2\ne 0 $ will be taken vanishingly small and finite for respectively zero and finite but small momentum transfer.
\end{widetext}
If one considers the various susceptibilities in the general case of two bands, the flow of the pair-vertex part $z^{\alpha\alpha}_{\text{CDW}}$ of intraband CDW$_\alpha$ susceptibility at ${\bm q}_\alpha $  is still governed by (\ref{z1bNadia_CDW}) derived previously in the separated band situation at finite phonon frequency, which is reproduced here with the notation of the present section:
\begin{align}
\label{z1bNadia_CDWb}
    \partial_\ell\ln z^{\alpha\alpha}_{\rm CDW} = &\, (-g_1^{\alpha\alpha;\alpha\alpha} + X\, g_2^{\alpha\alpha;\alpha\alpha}/2) \,\mathfrak{I}^{\alpha\alpha}_P.
    \end{align}
  
The finite phonon frequency will add vertex corrections to (\ref{zcdwpbp}) for the flow of the ${\bm q}_I$ interband pair vertex part $z^{\alpha\beta}_{\text{CDW}}$  for $\alpha\ne \beta$. These come from attractive interband couplings that reduce ${\bm q}_I$-CDW$_{\p\bp}$ correlations at low energy compared to the adiabatic situation, 
 \begin{align}
\label{}
   \partial_\ell & z^{\alpha\beta}_{\text{CDW}}  =  -g_1^{\beta\beta;\alpha\alpha} z^{\beta\alpha}_{\text{CDW}}  \mathfrak{I}_P^{\beta\alpha} -g_1^{\alpha\beta;\alpha\beta} z^{\alpha\beta}_{\text{CDW}}  \mathfrak{I}_P^{\alpha\beta}\cr 
   & + \frac{X}{2}\Big\{z^{\alpha\beta}_{\text{CDW}} g_2^{\beta\alpha;\alpha\beta}\mathfrak{I}_P^{\alpha\beta} +z^{\beta\alpha}_{\text{CDW}} g_2^{\beta\beta;\alpha\alpha}\mathfrak{I}_P^{\beta\alpha}\Big\}.
   \end{align}
We complete with the flow of $z_{\text{SC}}^{\alpha\alpha}$ for the superconducting susceptibility in the band $\alpha$,
\begin{align}
\partial_\ell\ln z^{\alpha\alpha}_{\rm SC} = &\, -\frac{X}{2} \Big\{  (g_1^{\alpha\alpha;\alpha\alpha} +g_2^{\alpha\alpha;\alpha\alpha}) \,\mathfrak{I}^{\alpha\alpha}_C \cr  & + (g_1^{\alpha\alpha;\beta\beta} +g_2^{\alpha\alpha;\beta\beta}) \,\mathfrak{I}^{\beta\beta}_C \Big\},
    \label{z1bNadia_SCb}
\end{align}
which  differs from the case of isolated bands in (\ref{z1bNadia_SC})   due to additional vertex corrections coming from attractive interband couplings that enhance SC$_\alpha$ correlations. 

\begin{figure*}
\includegraphics[width=1.0\linewidth]{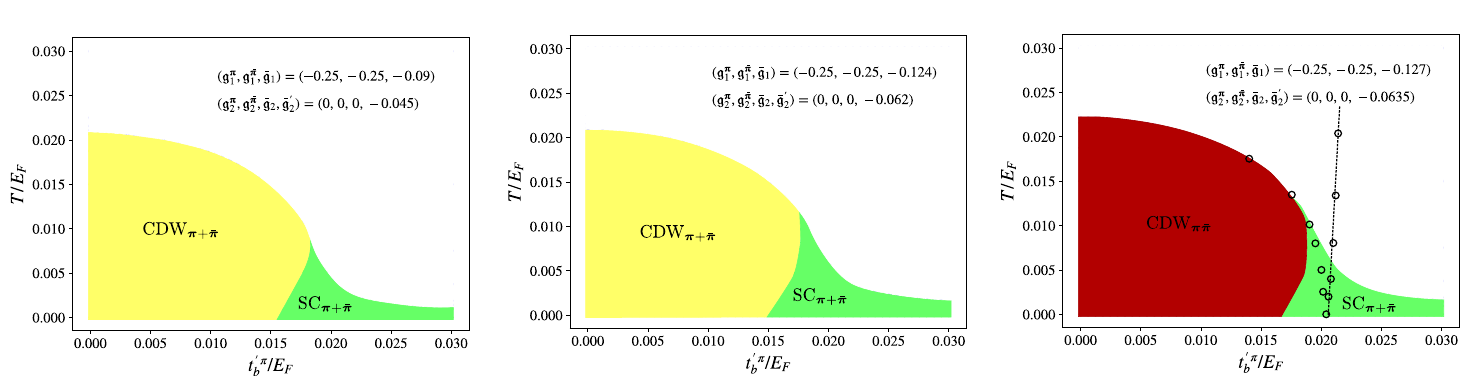}
\caption{Phase   for the coupled parallel bands ($\bar{v}=1$) at finite phonon frequency ($\bar{\omega}_D=0.033)$. (a) $  2|\bar{\mathfrak{g}}_1|<   |\mathfrak{g}^\alpha_1| $; (b) $  2|\bar{\mathfrak{g}}_1|\lesssim   |\mathfrak{g}^\alpha_1|$; (c) $2|\bar{\mathfrak{g}}_1|\gtrsim   |\mathfrak{g}^\alpha_1|$. The open circles correspond to the CDW energy scale $\Theta$ of the Curie-Weiss behavior, which emerges linearly from the unstable CDW$_{\p\bp}$ quantum critical point.   \label{DPhases_nb1_NA1}}
\end{figure*}

We begin with the study of possible phases of the model for totally parallel bands at $\bar{v}=1$. We consider equal intraband backscattering $\mathfrak{g}_1^{\p}=\mathfrak{g}_1^{\bp}$ and weaker interband analogs $\bar{\mathfrak{g}}_1$ such that $ 2|\bar{\mathfrak{g}}_1|<|\mathfrak{g}^\alpha_1| $, together with zero initial forward scattering $\bar{\mathfrak{g}}_2=\mathfrak{g}_2=0$ and  a 
finite but weak attractive interchain analog $\bar{\mathfrak{g}}'_2$. Following the integration of (\ref{g1IntraG}-\ref{z1bNadia_SCb}) up to $\ell\to \infty$, the $(T,t_b'^{\alpha})$ phase diagram of Fig.~\ref{DPhases_nb1_NA1}-(a) is then obtained. At low $t_b'^{\alpha}$, the CDW$_{\p}$ and CDW$_{\bp}$ phases with the same critical temperatures $T_\alpha$ are superimposed. As expected $T_\alpha$  slowly decreases  at finite  $t_b'^{\alpha}$ until a threshold of antinesting is reached where $T_\alpha $ drops rapidly. In contrast to the adiabatic situation of Fig.~\ref{DPhases_Inter_adia}, the CDW critical temperatures never reach zero and a singularity in both $z_{\mathrm{SC}}^{\p\p}$ and $z_{\mathrm{SC}}^{\bp\bp}$ takes over with the formation of a superconductivity dome that peaks at the threshold and is followed by a monotonic decay of $T_{c,\alpha}$  over the  remaining   $t_b'^{\alpha}$ interval.   

Increasing further the  interband coupling $|\bar{\mathfrak{g}}_1|$ alone, but with $ 2|\bar{\mathfrak{g}}_1|\lesssim |\mathfrak{g}^\alpha_1|$, the part of the phase diagram of Fig.~\ref{DPhases_nb1_NA1}-(b), with the superimposed CDW$_{\p}$ and CDW$_{\bp}$ phases remains essentially the same, while the superconductivity region  increases with a maximum of $T_{c,\alpha}$ at the dome peak, which is notably boosted as a direct consequence of the rise of interband coupling that enhances intraband Cooper pairing. The latter boost for superconductivity is essentially maintained when $|\bar{\mathfrak{g}}_1|$ is pushed further to reverse the inequality  $ 2|\bar{\mathfrak{g}}_1|\gtrsim  |\mathfrak{g}^\alpha_1|$, which favors the interband CDW $_{\p\bp}$ phase  at low  $t_b'^{\alpha}$, as shown in Fig.~\ref{DPhases_nb1_NA1}-(c). 

\begin{figure*}
\includegraphics[width=1.0\linewidth]{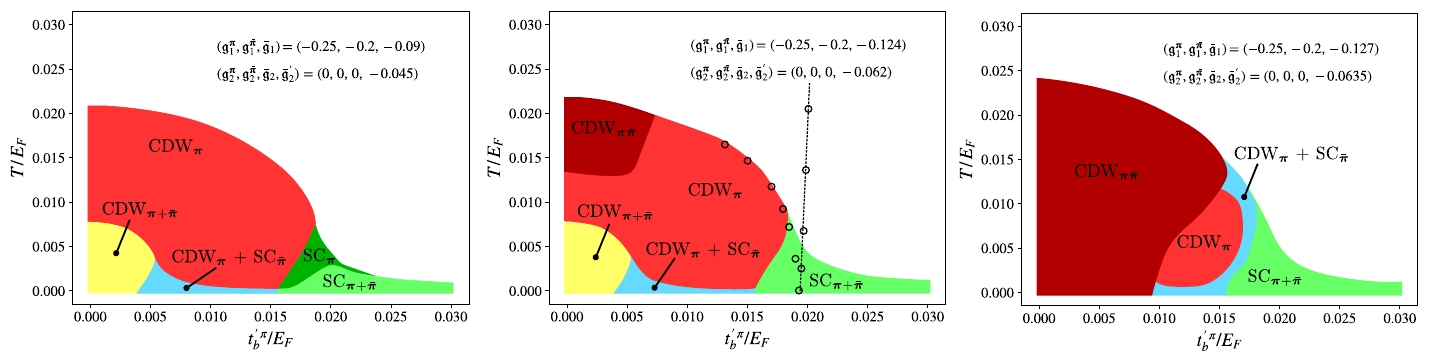}
\caption{Phase diagrams for the coupled asymmetric bands ($\bar{v}=1.25$) at finite phonon frequency ($\bar{\omega}_D=0.033)$. (a) $  2|\bar{\mathfrak{g}}_1|<   \mathrm{Max}\{|\mathfrak{g}^\alpha_1| \}$; (b) $  2|\bar{\mathfrak{g}}_1|\lesssim   \mathrm{Max}\{|\mathfrak{g}^\alpha_1| \}$; (c) $2|\bar{\mathfrak{g}}_1|\gtrsim   \mathrm{Max}\{|\mathfrak{g}^\alpha_1| \}$. The open circles correspond to the CDW energy scale $\Theta$ of the Curie-Weiss behavior, which emerges linearly from the unstable CDW$_{\p}$ quantum critical point.   \label{DPhases_nb08_NA1}}
\end{figure*}
We continue  the analysis  considering the influence of asymmetry between the  bands for a ratio $\bar{v}=1.25$ as shown in Fig.~\ref{DPhases_nb08_NA1}, along with the ratio of initial intraband values $\mathfrak{g}_1^{\p}/\mathfrak{g}_1^{\bp}=\bar{v}$ that is fixed accordingly. For relatively weak interband coupling that satisfies $ 2|\bar{\mathfrak{g}}_1|< \mathrm{Max}\{|\mathfrak{g}^\alpha_1| \}$, the phase diagram given in Fig.~\ref{DPhases_nb08_NA1}-(a) presents the two distinct phases CDW$_\p$ and CDW$_{\bp}$, in each band at low $t_b'^{\alpha}$, the former with the highest critical temperature, following the inequality $|\mathfrak{g}_1^{\p}|> |\mathfrak{g}_1^{\bp}|$ in the initial values. In this region of antinesting,  both phases superimpose at  sufficiently low temperature, whereas at a certain  threshold along the $t_b'^{\bp}$ axis,  a superconducting dome appears in the $\bp$ band, which is superimposed with a CDW$_\p$ phase in the other band. By cranking up  $t_b'^{\alpha}$ the CDW$_\p$ phase is in turn  weakened and evolves toward a superconductivity dome of the same band $\p$. However, a reentrant SC$_{\bp}$ phase becomes visible and superimposed on the SC$_{\p}$ at lower temperature. Compared with the results of separate bands in Fig.~\ref{fig_1bNadia}, the magnified size of the coexistence region  is  a consequence of the coupling between the bands. 

We now turn to the situation where  $|\bar{\mathfrak{g}}_1|$ is increased  to reach $2|\bar{\mathfrak{g}}_1|\lesssim \mathrm{Max}\{|\mathfrak{g}^\alpha_1| \}$ close to the separatrix. In the low $t_b'^{\bp}$ part of the phase diagram in Fig.~\ref{DPhases_nb08_NA1}-(b) a finite region of the CDW$_{\p\bp}$ interband phase emerges at low $t_b'^{\bp}$, to the loss of  the region of the  CDW$_\p$ phase compared to the previous case. Then both phases  become unstable, with the formation of an enlarged superconductivity dome with both SC$_{\p}$ and  SC$_{\bp}$ phases superimposed.

According to  Fig.~\ref{DPhases_nb08_NA1}-(c), when $|\bar{\mathfrak{g}}_1|$ increases further and crosses the separatrix with $2|\bar{\mathfrak{g}}_1|\gtrsim  \mathrm{Max}\{|\mathfrak{g}^\alpha_1| \}$, the interband CDW$_{\p\bp}$ phase imposed itself on the single band CDW$_\p$ and CDW$_{\bp}$ orders, although less rapidly  compared to the case of perfectly parallel bands  in Fig.~\ref{DPhases_nb1_NA1}, due to alteration of interband nesting at $\bar{v}\ne 1$. With respect to superconductivity, its enlargement continues, mainly due to the proximity to the remaining CDW$_\p$ phase with which  SC$_{\bp}$ of the other band can coexist. This region is followed at lower temperature by the coexistence of superconductivity in both bands.  Not  shown in the Figure is that for larger $|\bar{\mathfrak{g}}_1|$ where the CDW$_\p$ phase is completely suppressed to the benefit of the interband CDW$_{\p\bp}$ phase. This is followed by a superconductivity region  for both bands, but whose amplitude begins to weaken with a further increase of $|\bar{\mathfrak{g}}_1|$, a situation that becomes qualitatively similar to the one of Fig.~\ref{DPhases_nb1_NA1}-(c) for $\bar{v}=1$.  
\subsection{Quantum criticality}
\begin{figure*}
\includegraphics[width=1.0\linewidth]{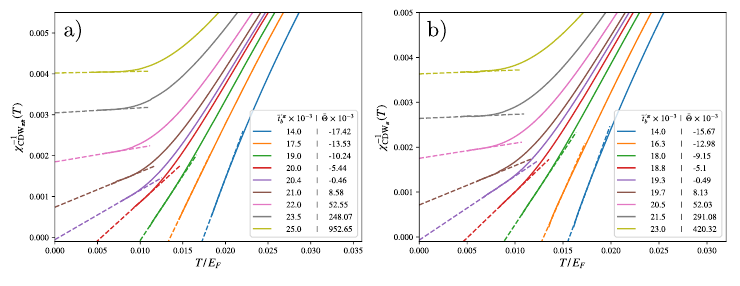}
\caption{The metallic phase inverse CDW susceptibility as a function of temperature and for different nesting alteration parameter $\bar{t}_b'^{\pi}={t}_b'^{\pi}/E_F$ close to the end boundary between CDW, SC. The dashed lines correspond the Curie-Weiss quantum critical behavior and $\bar{\Theta}= \Theta/E_F$ is a related characteristic  energy scale. (a): CDW$_{\p\bp}$ $-$ SC$_{\p+\bp}$ boundary of Fig.~\ref{DPhases_nb1_NA1}-(c); (b):  CDW$_\p$ $-$SC$_{\p+\bp}$ boundary of Fig.~\ref{DPhases_nb08_NA1}-(b). \label{QCP} }
\end{figure*}
From the above results, the dome of superconductivity that regularly originates from a dying out CDW  on the brink of a metallic state is a manifestation of  a quantum critical point that  in the end becomes unstable against the occurrence of superconductivity. Quantum fluctuations that derive from  a zero temperature critical point lying virtually at the foot of the dome are nevertheless expected to affect the normal state  at finite temperature. 
The  calculations of the present model reveal that these fluctuations leave an imprint on the temperature dependence of the CDW susceptibility, $\chi_{\mathrm{CDW}}(\bm{q}_{\gamma\beta})$, in a way similar to that obtained  previously in the context of an unstable quantum critical point of the spin-density wave state  \cite{Sedeki12,Bourbon09}.  

 This is illustrated in the inverse CDW susceptibility plots of Fig.~\ref{QCP}~(a) and (b), as a function of anitinesting for two examples of QCP connecting CDW, SC and normal states in the two-band situation, corresponding to  figures \ref{DPhases_nb1_NA1}-(c) and \ref{DPhases_nb08_NA1}-(b). These  plots exclude the area associated to the singularity of the transition.  On the CDW side of each phase diagram, but near the critical $t_b'^{*\alpha}$, $\chi^{-1}_{\mathrm{CDW}}(\bm{q}_{\gamma\beta})$ drops essentially linearly at high temperature, but as $t_b'^{\alpha}$ and antinesting becomes larger, it starts to bend, consistently with the drop of   $T_\alpha$ in the phase diagrams. At $t_b'^{*\alpha}$ where the system is superconducting, $\chi^{-1}_{\mathrm{CDW}}(\bm{q}_{\gamma\beta})$ evolves towards another linear $T$-regime that starts   above $T_{c,\alpha}$ and extrapolates to zero. This distinct linear regime can be well described by a Curie-Weiss temperature dependence for the susceptibility, 
 \begin{equation}
   \chi_{\mathrm{CDW}}(\bm{q}_{\gamma\beta})\approx \frac{C}{T+ \Theta} , 
 \end{equation}
 where $C$ is a constant and  $\Theta $ is a characteristic  energy scale of CDW order at $ \bm{q}_{\gamma\beta}$.  $\Theta$ goes from negative values, where it essentially merges with the critical temperature or the CDW gap   at low $t_b'^{\alpha}$, and evolves to zero at $t_b'^{*\alpha}$ under the superconducting dome to finally reach positive values in the presence of a superconducting fixed point for $t_b'^{\alpha} \ge t_b'^{*\alpha}$.  According to the results of Figure~\ref{QCP}, $\Theta\sim (t_b'^{\alpha}-t_b'^{*\alpha})^{\theta}$ rises from zero at the QCP with an exponent  $\theta= 1$, that is, linearly at  $t_b'^{\alpha}\ge t_b'^{*\alpha}$, as illustrated  in Figures -\ref{DPhases_nb1_NA1}-(c) and \ref{DPhases_nb08_NA1}-(b). This is in agreement  with the quantum scaling argument for $\theta = z\nu$, and the values $z=2$ and $\nu =1/2$ for  the exponents of the dynamics of correlations and their characteristic   length [$\xi \sim (T+ \Theta)^{-\nu}$] at the one-loop level of the RG. 

 It follows that, although no  longer singular, the growth of  $\xi_{\alpha\beta}$ and then of $\chi_{\mathrm{CDW}}(\bm{q}_{\gamma\beta})$ is    still present at $t_b'^{\alpha}\ge t_b'^{*\alpha}$, which is considered  a signature of  the quantum critical domain. Actually, this growth is directly related to the   positive quantum interference between the singular Cooper scattering channel and the weak but regular electron-hole pairing one; it  progressively weakens as $t_b'^{\alpha}$ moves away from the QCP and where $\Theta$ increases rapidly.   
\subsection{Isotope effect}
\begin{figure*}
\includegraphics[width=1.0\linewidth]{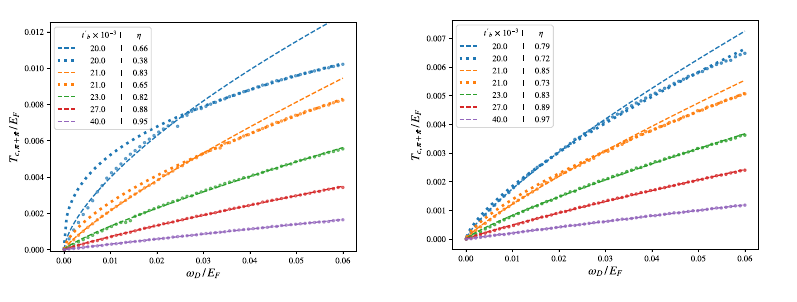}
\caption{Variation of the critical superconducting temperature $T_{c,\p+\bp}$  as a function of the reduced phonon frequency $\omega_D/E_F$ and different deviations $\bar{t}_b'^{\p}= t_b'^{\p}/E_F$. From the results of    Figure~\ref{DPhases_nb1_NA1}-(c) in (a)  and  Figure~\ref{DPhases_nb08_NA1}-(b) in (b). The dashed lines are power law fits $T_{c,\p+\bp}\sim \omega_D^\eta$. \label{Isotope}}
\end{figure*}

 The above RG results show that the critical temperature for the onset of superconductivity in a given band $\alpha$ is magnified by  quantum critical CDW fluctuations, compared to the bare limit of  BCS theory,  $T^0_{c,\alpha}=1.14\omega_D e^{-2/|\mathfrak{g}^\alpha_1|}$, obtained with the initial unrenormalized coupling $\mathfrak{g}^\alpha_1$.  This can be further quantified by looking at the isotope effect known to link the  variation  of $T_{c,\alpha} $ in $\omega_D$  ($\propto 1/\sqrt{M} $).  The influence of  quantum   fluctuations on the effective attraction between electrons that is shaping at low energy will reflect the departure from the linear dependence $ T_{c,\alpha} \propto \omega_D $ predicted by the  BCS theory. 

This is illustrated in Fig.~\ref{Isotope}-(a) and (b)  where the variation of the superconducting $T_{c,\alpha}$ with $\omega_D$ is plotted for $t_b'^{\alpha}\ge t_b'^{*\alpha}$ in 
coupled two-band situations corresponding to the initial conditions of Figures~\ref{DPhases_nb1_NA1}-(c) and \ref{DPhases_nb08_NA1}-(b).  
In the first case shown in  Figure~\ref{Isotope}-(a) for perfectly parallel bands, the deviations from  the  BCS isotope effect are the strongest for $t_b'^\p\approx t_b'^{*\p}$, namely  in  the region close to the QCP.  The deviations can be expressed in the form of power laws $T_{c,\p+\bp}\sim \omega_D^\eta$ for   the critical temperature. At approximately  $t_b'^{*\p}$ these can actually be well fitted by a sequence of two power laws as a function of $\omega_D$, one with $\eta\sim 0.7$ at relatively low phonon frequency and  which is followed by another with the smaller  $\eta\sim 0.4$ at higher frequency. This indicates two different regimes of retardation effects  on the renormalization of the coupling constants,  each differing from the BCS result $\eta=1$. Moving away from $t_b'^{*\p}$, the sequence is still present, but with a weaker difference in the values of $\eta$, both of which increase consistently with the decay of quantum fluctuations. This continues further above $t_b'^{*\p}$ where a single power law is found throughout the whole frequency range, with  a value of $\eta $ that continues to increase. As shown in the Figure, with larger $t_b'^{\p}$ the results  continuously evolve towards the linear BCS limit as the influence of fluctuations on the coupling renormalization becomes much weaker.  

A qualitatively similar   isotope effect is found for the results of Fig.~\ref{DPhases_nb08_NA1}-(b) for non parallel bands, as shown in Fig.~\ref{Isotope}-(b). Only the exponents are   slightly larger. 
\section{Summary and concluding remarks}
\label{conclusion}

 The observation of superconductivity in proximity  of charge-density wave order in a variety of low-dimensional materials brought into focus  the possible mechanisms that can be responsible for this closeness and how a multiple component electronic structure can be a positive factor in its development.  In this  work,  we addressed this problem with the aid of a perturbative renormalization group method that can incorporate a  two-band  quasi-1D electron model and retarded phonon-mediated interactions. The multiplicity of charge-density  wave and superconducting ordered  states   can be  established, and  their competing character can be analyzed in calculated phase diagrams. These are obtained at finite temperature  and from the tunable nesting frustration which acts as a parameter simulating the important role played by applied pressure  in the electronic structure of  concrete materials. 

 The results for the limiting case of isolated bands have been  reviewed in the infinite retardation  or adiabatic limit, where the renormalization group method turns out to be equivalent to the mean-field theory prediction of critical temperatures for   different  charge-density wave ordered states. These states can be superimposed, but each is invariably critically suppressed at some finite threshold of antinesting. Beyond the adiabatic limit  when phonon frequency is non zero, there is a finite electron energy domain where phonon-mediated interactions become essentially non-retarded. This initiates Cooper pairing for electrons, which, together with the above suppression of charge-density wave states,  yields quantum critical conditions for a dome of superconductivity  to issue from in each band.

In the general case where the electrons of the two bands interact through phonon-mediated  couplings, the adiabatic limit sees, besides the  modulated state of each band, the development of an interband  charge-density wave phase that can intrude into  the phase diagram at low antinesting and adds to the multiplicity of modulated states. It can even dominate whenever the strength of interband interactions reaches some characteristic value with respect to their intraband counterparts.  
When the phonon frequency is finite and Cooper pairing  quantum mechanically interferes with electron-hole pairing tied to the variety of charge-density wave fluctuations, a superconductivity dome can be stabilized in one or both  bands. The dome peaks on the edge of the critical decline of parent charge-density wave phases along the antinesting axis. The coupling between the bands is found to act as a notable source of enhancement of  superconducting critical temperatures compared to the single band results. This likely points to the multiple component of the electronic spectrum as a contributing factor to the observability of superconductivity in real multicomponent electron systems. 

The results developed in the present work are also congruent with a range of features found in several low dimensional charge-density wave superconductors in which the coupling of electrons to phonons  proves to be  important  in a multicomponent electronic structure. This is the case of the dome of stable superconductivity, which under pressure  often shows up on the edge of a falling out charge ordered state, itself with different  components that can be superimposed. Such a sequence of phases is particularly manifest, for instance, in  multiband quasi-1D organics and transition metal trichalcogenides inorganics systems \cite{Kaddour14,Brossard89,Monceau12,Pouget24}.  

The confirmation of quantum critical domain of fluctuations predicted to be present as precursors to the dome of superconductivity is also of importance for the relevance of the present model in concrete situations. The  fingerprints of quantum criticality should be observable by scattering techniques that can probe finite wave vector correlations coming from charge modulations \cite{Pouget24}. The same quantum correlations can be responsible for the anomalous temperature dependence seen in the electrical resistivity and  Seebeck coefficient \cite{Monceau12}. Although difficult to realize in practice, the isotope effect discussed in this work can act as a relevant tool to probe the influence of quantum fluctuations on the critical temperature of superconductivity.     

\begin{acknowledgments}
C. B. thanks the National Science and Engineering Research Council of Canada (NSERC),the Regroupement Québécois des Matériaux de Pointe (RQMP) and the Institut Quantique of Université de Sherbrooke for financial support. The authors thank  Lucas Désoppi and Émile Larouche for useful comments on the numerical solution of renormalization group equations. 
\end{acknowledgments}

\appendix
\section{Peierls loop}
In this appendix, we proceed to the evaluation of the coefficient of the Peierls closed loop of Fig.~\ref{loops} for the flow equations. When evaluated in the static limit at $\bar{q}_{\gamma'\beta'}= ({\bm q}_{\gamma'\beta'},0) $, one finds, after summation over Matsubara frequencies, 
\begin{align}
 \mathfrak{I}_{\mathrm{P}}^{\gamma'\beta'} = & - \frac{2T}{N_b}\sqrt{\frac{v_F^{\beta'}}{v_F^{\gamma'}}}\sum_{\omega_{\bar{n}},\bar{k}_b} \partial_\ell\intslashD {d\bar{{E}}^{\gamma'}} G^{\beta'}_+(\bar{k} +\bar{q}_{\gamma'\beta'})G^{\gamma'}_-(\bar{k}) \cr
 = & \frac{1}{N_b}\sum_{\bar{k}_b}\sqrt{\frac{v_F^{\beta'}}{v_F^{\gamma'}}}\partial_\ell\intslashD {d\bar{E}_-^{\gamma'}} \frac{n[\bar{E}_{-,{\bm k}}^{\gamma'}] -n[\bar{E}_{+,{\bm k}+ {\bm q}_{\gamma'\beta'}}^{\beta'}] }{\bar{E}_{+,{\bm k}+ {\bm q}_{\gamma'\beta'}}^{\beta'} - \bar{E}_{-,{\bm k}}^{\gamma'}}
 \label{}
\end{align}
where $n[x]=(e^{\beta x} +1)^{-1}= \frac{1}{2}[1-\tanh \beta x/2]$ is the Fermi-Dirac distribution. For the best nesting vector between the $\gamma'$ and $\beta'$ bands, when ${\bm q}_{\gamma'\beta'}=(k_F^{\gamma'}+k_F^{\beta'},\pi)$, the electron and hole spectrum is connected by the relation,
$$
\bar{E}_{+,{\bm k}+ {\bm q}_{\gamma'\beta'}}^{\beta'} = - \frac{v_F^{\beta'}}{v_F^{\gamma'}} [\bar{E}_{-,{\bm k}}^{\gamma'} + 4 t_b'^{\gamma'} \cos 2\bar{k}_b ]
$$ 
where the $t_b'^{\gamma'}$ term displays deviations to perfect electron-hole symmetry (antinesting). From the shell energy integral, ${\intslash d\bar{E}^{\gamma'} = \big(\int_{-E_0(\ell)/2}^{-E_0(\ell+d\ell)/2} + \int_{E_0(\ell)/2}^{E_0(\ell+d\ell)/2}\big) d\bar{E}^{\gamma'}}$, where $E_0(\ell)= 2E_F e^{-\ell}$, one obtains the expression of the Peierls loop coefficient dependent on  temperature,  antinesting and band asymmetry 
\begin{widetext}
\begin{align}
\mathfrak{I}_{\mathrm{P}}^{\gamma'\beta'}=  \frac{\sqrt{v_F^{\beta'}v_F^{\gamma'}}}{v_F^{\gamma'}+ v_F^{\beta'}}\left\langle  \frac{\tanh[e^{-\ell}/2\bar{T}] + \tanh[(v_F^{\beta'}/2v_F^{\gamma'}\bar{T})( e^{-\ell} - 4 \bar{t}_b'^{\gamma'} \cos 2k_b )]}{1+ 4 e^\ell(1+ v_F^{\gamma'}/v_F^{\beta'})^{-1} \bar{t}_b'^{\gamma'} \cos 2k_b}\right\rangle_{\bar{k}_b},
\label{IPloop}
\end{align}
where $\langle \ldots\rangle_{\bar{k}_b}$ is the average over the transverse wave vector $\bar{k}_b$ and $\bar{T}= T/E_F$, $\bar{t}_b'^{\gamma'}=t_b'^{\gamma'}/E_F $. When the  bands are distinct but perfectly parallel, with $(v_F^{\beta'},t_b^{\beta'},t_b'^{\beta'})=(v_F^{\gamma'},t_b^{\gamma'},t_b'^{\gamma'})$ or $\bar{v}=1$, the expression reduces to 
\begin{align}
     \mathfrak{I}^{\gamma'\beta'}_P = 
   \frac{1}{2} & \Big\langle \frac{\tanh [e^{-\ell}/2\bar{T}] + \tanh [e^{-\ell}/2\bar{T} - 2({\bar{t}'_b}\!^{\gamma'}/\bar{T})\cos 2\bar{k}_b]}{{  1 + 2{\bar{t}'_b}\!^{\gamma'} e^{\ell} \cos 2\bar{k}_b}}   \Big\rangle_{\bar{k}_b}.
\label{IPalpha}
\end{align}
The latter expression coincides to  the case of shell intraband Peierls loop, when $\gamma'=\beta'=\alpha$ and 
\begin{align}
     \mathfrak{I}^{\alpha\alpha}_P = 
   \frac{1}{2} & \Big\langle \frac{\tanh [e^{-\ell}/2\bar{T}] + \tanh [e^{-\ell}/2\bar{T} - 2({\bar{t}'_b}\!^\alpha/\bar{T})\cos 2\bar{k}_b]}{{  1 + 2{\bar{t}'_b}\!^\alpha e^{\ell} \cos 2\bar{k}_b}}   \Big\rangle_{\bar{k}_b}.
\label{IPalpha}
\end{align} 
and where $\alpha=\p,\bp$.  The above expressions are involved in the Peierls scattering channel of the RG equations of Sec.~\ref{RGEqs}. 
 \end{widetext}

 \section{Determination of   boundaries of phase diagrams at finite temperature }
 \label{Boundary}
 \begin{figure}[b]
\includegraphics[width=\linewidth]{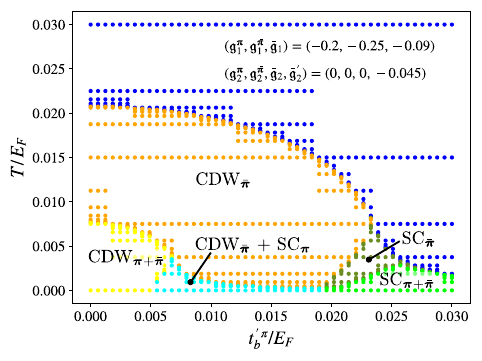} 
\caption{\label{Dicho} Raw RG data for the dichotomy determination of a phase diagram here for the coupled asymmetric bands ($\bar{v}=0.8$) at finite phonon frequency ($\bar{\omega}_D=0.033)$ and $  2|\bar{\mathfrak{g}}_1|<   \mathrm{Max}\{|\mathfrak{g}^\alpha_1| \}$.}
\end{figure}

 For the determination of phase boundaries from  the solution of the RG flow equations, we have adapted the dichotomy approach used in ~\cite{Desoppi24}, to finite temperature conditions.  This procedure  is illustrated in  Figure~\ref{Dicho} where  the raw numerical data of the phase diagram for the model are presented in the two-band situation at finite phonon frequency. The case complements Fig.~\ref{DPhases_nb08_NA1}-(a) with $\bar{v}< 1$ which inverted the initial conditions on $\mathfrak{g}_1^\p$ and  $\mathfrak{g}_1^{\bp}$. Every $\p$ (resp. $\bp$) phase of Fig.~\ref{DPhases_nb08_NA1}-(a) then turns to a $\bp$ (resp. $\p$) phase. 

 The dichotomy procedure is carried out by discrete steps along the vertical temperature axis within the interval $[t'^{\p}_b=0,T/E_F=0.03]$, here chosen to include all possible phases for the given initial conditions on the couplings. The first run of RG flow is carried out on the top left at  the (0, 0.03) blue point, where no singularity is encountered which corresponds to a non-ordered metallic state. The next run is carried out at the middle point (0, 0,015) of the interval, where a singularity is found for CDW$_{\bp}$ order  (orange point). The next RG flow is obtained above in the middle of the top interval at (0, 0,022), which brings the system  back  into the unordered state. This procedure is continued until the boundary CDW$_{\bp}$-Metal is reached. The same procedure is applied in the lower part of the interval [0.0.015], which allows identifying a boundary between the CDW$_{\bp}$ and the superimposed CDW$_{\p+\bp}$ phase (yellow points). The dichotomy is then performed on a grid of 30 points along the   
 $t'^{\p}_b$ axis, which allows us to obtain the complete phase diagram of Figure~\ref{Dicho}. An interesting feature that emerges from the procedure conducted at finite temperature is the possibility of incursion of  superconducting phases within the CDW ones, which leads to dome of stability for superconductivity.

\bibliography{aipsamp}

\end{document}